\documentclass{article}

\usepackage{amsmath}
\usepackage{amssymb}

\usepackage{hyperref}
\usepackage[capitalize,noabbrev]{cleveref}

\usepackage{xcolor}

\usepackage{xspace}

\usepackage{multirow}
\usepackage{makecell}

\usepackage{enumitem}

\definecolor{olivegreen}{rgb}{0, 0.6, 0}
\definecolor{blue(ncs)}{rgb}{0.0, 0.53, 0.74}

\usepackage{pifont}

\usepackage{pgfplots}
\pgfplotsset{compat=1.18}

\definecolor{pastelblue}{RGB}{174, 198, 207}
\definecolor{pastelred}{RGB}{255, 179, 186}
\definecolor{pastelgreen}{RGB}{153, 204, 153}

\usepackage{colortbl}

\usepackage{microtype}
\usepackage{graphicx}
\usepackage{booktabs}

\usepackage[accepted]{others/mlsys2025}

\newcommand{\thiswork}{GriNNder\xspace}

\newcommand{\genstructure}{structured storage offloading\xspace}
\newcommand{\Genstructure}{Structured storage offloading\xspace}
\newcommand{\GenStructure}{Structured Storage Offloading\xspace}

\newcommand{\ourstructure}{partition-wise graph caching\xspace}
\newcommand{\Ourstructure}{Partition-wise graph caching\xspace}
\newcommand{\OurStructure}{Partition-Wise Graph Caching\xspace}

\newcommand{\ourgrad}{grad-engine activation regathering\xspace}
\newcommand{\Ourgrad}{Grad-engine activation regathering\xspace}
\newcommand{\OurGrad}{Grad-Engine Activation Regathering\xspace}

\newcommand{\ourpart}{switching-aware partitioning\xspace}
\newcommand{\Ourpart}{Switching-aware partitioning\xspace}
\newcommand{\OurPart}{Switching-Aware Partitioning\xspace}

\newcommand{\micro}{MFG-based\xspace}

\newcommand{\ourpolicy}{intra-layer partition-wise caching\xspace}

\usepackage{tikz}
\usetikzlibrary{shapes.geometric}

\newcommand*\circled[1]{\tikz[baseline=(char.base)]{
            \node[shape=circle,draw,inner sep=0.4pt, fill=white, text=black] (char) {#1};}}

\newcommand*\bcircled[1]{\tikz[baseline=(char.base)]{
            \node[shape=circle,inner sep=0.4pt, fill=black, text=white] (char) {#1};}}

\newcommand{\adjustdescender}[1]{%
  \ifx#1g \raisebox{0.3ex}{g}%
  \else\ifx#1y \raisebox{0.3ex}{y}%
  \else\ifx#1p \raisebox{0.3ex}{p}%
  \else\ifx#1j \raisebox{0.3ex}{j}%
  \else\ifx#1q \raisebox{0.3ex}{q}%
  \else #1%
  \fi\fi\fi\fi\fi
}

\newcommand*\acircled[1]{\tikz[baseline=(char.base)]{
            \node[shape=circle,draw, fill=white, text=black, inner sep=0pt, minimum size=1em] (char)  {\adjustdescender{#1}}; }}
            
\newcommand{\PentagonText}[1]{%
    \tikz[baseline=(char.base)]{
        \node[shape=regular polygon, regular polygon sides=5, draw, fill=white, text=black, inner sep=0.6pt, minimum size=0.9em, text height=1ex, text depth=0.25ex, anchor=center] (char)  {\raisebox{-0.15ex}{#1}};
    }%
}

\newcommand{\HexagonText}[1]{%
    \tikz[baseline=(char.base)]{
        \node[shape=regular polygon, regular polygon sides=6, draw, fill=white, text=black, inner sep=0.6pt, minimum size=0.9em, text height=1ex, text depth=0.25ex, anchor=center] (char)  {\raisebox{-0.15ex}{#1}};
    }%
}

\newcommand{\naive}{naïve\xspace}
\newcommand{\Naive}{Naïve\xspace}
\newcommand{\naively}{naïvely\xspace}

\newcommand{\mr}[2]{\multirow{#1}{*}{#2}}

\mlsystitlerunning{\thiswork: Breaking the Memory Capacity Wall in Full-Graph GNN Training with Storage Offloading}

\begin{document}

\twocolumn[
\mlsystitle{\thiswork: Breaking the Memory Capacity Wall in\\ Full-Graph GNN Training with Storage Offloading}

\begin{mlsysauthorlist}
\mlsysauthor{Jaeyong Song}{snu}
\mlsysauthor{Seongyeon Park}{snu}
\mlsysauthor{Hongsun Jang}{snu}
\mlsysauthor{Jaewon Jung}{snu}
\mlsysauthor{Hunseong Lim}{snu}
\mlsysauthor{Junguk Hong}{snu}
\mlsysauthor{Jinho Lee}{snu}
\end{mlsysauthorlist}

\begin{center}
\vspace{-1mm}
\url{https://github.com/AIS-SNU/GriNNder} 
\vspace{-8mm}
\end{center}

\mlsysaffiliation{snu}{Department of Electrical and Computer Engineering, Seoul National University, Seoul, South Korea}

\mlsyscorrespondingauthor{Jinho Lee}{leejinho@snu.ac.kr}

\mlsyskeywords{Graph Neural Networks, GNN Training, Storage Offloading, Training Frameworks}

\vskip 0.3in

\begin{abstract}
Full-graph training of graph neural networks (GNNs) is widely used as it enables direct validation of algorithmic improvements by preserving complete neighborhood information. 
However, it typically requires multiple GPUs or servers, incurring substantial hardware and inter-device communication costs.
While existing single-server methods reduce infrastructure requirements, they remain constrained by GPU and host memory capacity as graph sizes increase.
To address this limitation, we introduce \emph{\thiswork}, which is the first work to leverage storage devices to enable full-graph training even with limited memory.
Because modern NVMe SSDs offer multi-terabyte capacities and bandwidths exceeding 10 GB/s, they provide an appealing option when memory resources are scarce.
Yet, directly applying storage-based methods from other domains fails to address the unique access patterns and data dependencies in full-graph GNN training.
\thiswork tackles these challenges by \emph{\genstructure (SSO)}, a framework that manages the GPU-host-storage hierarchy through coordinated \emph{cache}, \emph{(re)gather}, and \emph{bypass} mechanisms.
To realize the framework, we devise (i) a partition-wise caching strategy for host memory that exploits the observation on cross-partition dependencies, (ii) a regathering strategy for gradient computation that eliminates redundant storage operations, and (iii) a lightweight partitioning scheme that mitigates the memory requirements of existing graph partitioners.
In experiments performed over various models and datasets, \thiswork achieves up to 9.78$\times$ speedup over state-of-the-art baselines and throughput comparable to distributed systems, enabling previously infeasible large-scale full-graph training even on a single GPU.
\end{abstract}

]

\printAffiliationsAndNotice{}

\section{Introduction}
\label{sec:intro}

Graph neural networks (GNNs) have emerged as essential tools for learning from graph-structured data, targeting social networks~\cite{social_gnn_survey}, molecular interactions~\cite{protein_gnn}, and computer vision~\cite{image_gnn_survey}.
As graphs can capture arbitrary relationships among entities, GNNs hold broad potential across diverse domains.

Among GNN training paradigms, full-graph training~\cite{pipegcn, bns_gcn, roc, gnnautoscale, lmc, sancus, granndis} processes the entire graph per iteration, avoiding information loss.
This provides high accuracy and theoretical guarantees, simplifying algorithmic validation.
Our survey on recent GNN publications (\cref{app:survey}) reveals that many of them select full-graph training for these advantages, especially when the accuracy upper bound is unknown for new tasks or methods.

However, full-graph training requires storing all node activations and gradients across all GNN layers in memory, easily exceeding modern GPU capacity.
Some single-server methods~\cite{betty, hongtu} have been proposed, but remain fundamentally limited by GPU or host memory capacity for large graphs (\cref{app:prev_works}).
While distributing workload across multiple GPUs is possible, this introduces significant hardware cost and inter-device communication overhead, often leading to poor scalability~\cite{sancus, pipegcn}.

These hardware-imposed limitations constrain researchers from flexibly designing and validating algorithms.
Many studies in our survey (\cref{app:survey}) either co-design complicated memory-saving algorithms (e.g., sampling~\cite{graphsage}) to fit data in memory \cite{full_vs_mini}, or report out-of-memory failures with large graphs rather than scaling to distributed environments.

To address this, we introduce \thiswork, the first framework that breaks through GPU and host memory limitations by leveraging storage as an additional memory hierarchy tier.
Modern NVMe SSDs provide TB-scale storage and exceed 10 GB/s in bandwidth, making them practical for storing large volumes of intermediate training data.
However, no prior full-graph training system has effectively exploited this storage tier---not because of fundamental hardware constraints, but because of the rigidity of existing frameworks.

One might expect that storage-based methods from other domains could alleviate GPU and host memory limits.
However, these solutions cannot be directly applied to full-graph GNN training.
LLM frameworks~\cite{zero_infinity, flexgen} mainly offload model weights, but the weights' memory is negligible in GNNs because parameters are shared across all vertices.
Similarly, mini-batch-based GNN training systems with storage~\cite{ginex, marius, diskgnn, gnndrive} leverage storage only to cache input features rather than intermediate activations and gradients.
Extending them to full-graph settings (so-called micro-batch training~\cite{betty}) inherits the same constraints while still suffering from GPU out-of-memory due to neighbor explosion (\cref{app:extension_limit}).

When employing storage for full-graph GNN training, three key system-level challenges emerge:
\begin{enumerate}[leftmargin=*, itemsep=0pt, topsep=0pt]
\item \emph{Storage I/O Bottlenecks}: 
Despite improved NVMe SSD bandwidth, storage remains far slower than host memory and suffers from inefficient I/O due to fixed page granularity and random access patterns.
\item \emph{Data Amplification}: 
Existing frameworks~\cite{pytorch, pyg, hongtu} utilize activation snapshots to enable sequential accesses.
However, this becomes impractical when used with storage, substantially inflating both memory usage and I/O traffic.
\item \emph{Memory-Hungry Partitioning}: 
Full-graph training requires partitioning the graph until memory requirements fit GPU capacity.
However, the standard partitioner~\cite{metis, parmetis, mtmetis} used in prior approaches~\cite{betty, hongtu} often exceeds host memory limits during partitioning itself, necessitating a separate large-memory server that may harm practicality.
\end{enumerate}

To address these challenges, we propose \emph{\thiswork}, which introduces \textit{\genstructure (\textbf{SSO})}, a general framework for managing the GPU-host-storage memory hierarchy in full-graph GNN training.
SSO orchestrates three coordinated mechanisms---\textit{cache}, \textit{(re)gather}, and \textit{bypass}---to enable efficient storage-aware training.
This strategy is realized through the following specialized methods:
\begin{itemize}[leftmargin=*, itemsep=0pt, topsep=0pt]
    \item \Ourstructure: 
    We observe that cross-partition dependencies follow a power-law distribution, similar to the degree distributions of real-world graphs.
    Exploiting this, we design partition-wise caching that uses host memory as an efficient cache with optimized I/O policies, minimizing inefficient storage access.
    \item \Ourgrad: 
    A regathering strategy for the automatic gradient computation engine. It eliminates inefficient activation snapshotting in existing offloading solutions, thereby minimizing redundant data storage and movement.
    \item \Ourpart: 
    A lightweight, memory-efficient partitioning algorithm specifically designed for host-memory-limited environments, avoiding the high memory footprint of standard graph partitioners.
\end{itemize}

We implement \thiswork as \texttt{PyGriNNder}, enabling users to leverage existing PyTorch Geometric~\cite{pyg} code by simple inheritance.
Notably, \thiswork does not modify the training algorithm itself, ensuring seamless migration without risk of accuracy degradation.
Our experiments demonstrate that \thiswork achieves up to 9.78$\times$ speedup over the state-of-the-art and throughput comparable to distributed baselines, enabling previously infeasible large-scale full-graph training even on a single GPU.

\begin{figure}[t]
    \centering
    \includegraphics[width=\columnwidth]{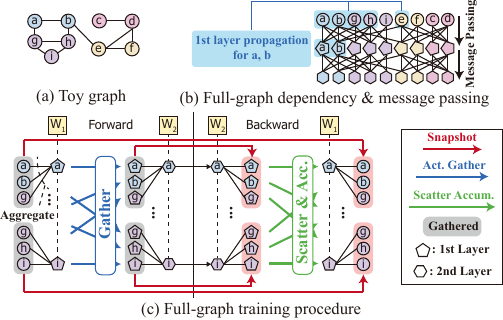}
    \vspace{-5mm}
    \caption{
    Full-graph training procedure with a two-layer GNN.
    }
    \label{fig:background}
\end{figure}

\section{Background: Full-Graph GNN Training}
\label{sec:background}

Full-graph GNN training~\cite{roc, cagnet, sancus, bns_gcn, pipegcn} processes the entire graph in each training iteration without sampling, propagating information through message passing across multiple layers.
Unlike subgraph training (e.g., mini-batch~\cite{graphsage}), it uses all edge connections, preserving complete neighborhood information.
This approach simplifies algorithmic validation but requires storing intermediate activations/gradients for all $|V|$ vertices across all $|L|$ layers simultaneously with $|H|$ hidden dimensions---often TBs for large graphs.

\cref{fig:background} illustrates full-graph training using a two-layer GNN on the toy graph in \cref{fig:background}a.
\cref{fig:background}b shows the two-layer dependency structure derived from this topology.
Starting from input features (denoted with circled vertex IDs), features are propagated via \emph{message passing} to destination vertices in the intermediate layer (e.g., \acircled{a}, \acircled{b}, \acircled{g}, \acircled{h}, \acircled{e} $\xrightarrow{}$ \PentagonText{a}, \PentagonText{b}, shaded blue).
The second layer applies the same process using these intermediate features to produce the final output embeddings.

\cref{fig:background}c illustrates the layer-by-layer training procedure.
\textbf{Forward Pass.}
To compute an output feature vector, features from source vertices in the previous layer must be \emph{aggregated} (e.g., averaged).
For example, vertex feature \PentagonText{a} depends on \acircled{a}, \acircled{b}, and \acircled{g}, including an implicit self-loop.  
After aggregation, the features are multiplied with the shared weight matrix (e.g., $W_1$), then processed through operations such as normalization and activation to produce the layer's output features (denoted by \PentagonText{}).
For the next layer, these output features are \emph{gathered} to create inputs for aggregation following the same dependency structure (blue arrows).
The gathered activations are saved as \emph{snapshots} in GPU or host memory for use in the backward pass.

\textbf{Backward Pass.}
During backpropagation, the dependency flow is inverted.
The gradient of vertex feature \HexagonText{a} is propagated to \PentagonText{a}, \PentagonText{b}, and \PentagonText{g} to compute their gradients.
This requires loading the previously saved snapshots (red arrows), then \emph{scatter-accumulating} the computed gradients to the vertices of the previous layer (green arrows).

When workloads fit in GPU memory, this procedure enables fast training through massive parallelism and high memory bandwidth. 
However, severe capacity pressure arises: the entire intermediate data, including activations/gradients, must fit within GPU memory.
A straightforward solution is distributed training~\cite{cagnet, sancus}, but this incurs substantial hardware costs and inter-device communication overhead (often 80--98\%, see \cref{app:prev_works}).

To mitigate GPU memory constraints, single-server methods~\cite{betty, hongtu} have been recently proposed.
However, they remain limited by GPU and host memory capacity and require memory-hungry partitioning operations that consume hundreds of GBs.
We provide a detailed analysis of these limitations in \cref{app:prev_works}.

\begin{figure}[t]
    \centering
    \includegraphics[width=.9\columnwidth]{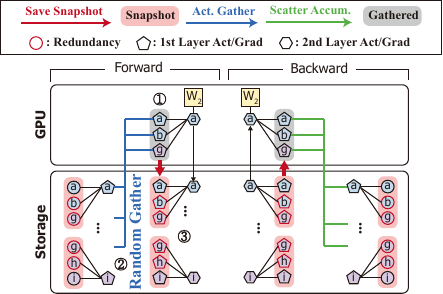}
    \vspace{-2mm}
    \caption{\Naive storage extension of full-graph training.}
    \label{fig:naive_ext}
    \vspace{-4mm}
\end{figure}

\textbf{\Naive Storage Employment.}
Given full-graph training as described in \cref{fig:background}, a straightforward method would place the small weights (and gradients) on the GPU and large activations (and gradients) on storage.
\cref{fig:naive_ext} illustrates an example procedure for processing a single vertex, \HexagonText{a}.
Since the neighbors of \HexagonText{a} (\PentagonText{a}, \PentagonText{b}, and \PentagonText{g}) fit within GPU memory, training can proceed.
However, this approach yields sub-optimal performance for three main reasons:
\circled{1} Ensuring gathered neighbor features fit within GPU memory is challenging due to power-law degree distributions.
Memory requirements per partition vary dramatically, making GPU capacity violations difficult to avoid without memory-aware partitioning.
\circled{2} Gathering feature vectors requires random reads from storage.
Since storage devices operate at page granularity (e.g., 16 KiB for NVMe SSDs), random access leads to severe read amplification and bandwidth saturation.
\circled{3} While existing snapshot features in PyTorch~\cite{pytorch} and prior methods~\cite{hongtu} enable sequential access, they introduce significant \emph{redundancy}, inflating write traffic.
For instance, \PentagonText{g} appears redundantly in snapshots of all neighboring vertices—\HexagonText{a}, \HexagonText{h}, and \HexagonText{i}.

\section{\GenStructure: Cache-(Re)Gather-Bypass}
\label{sec:overview}

\begin{figure}[t]  
    \centering
    \includegraphics[width=\columnwidth]{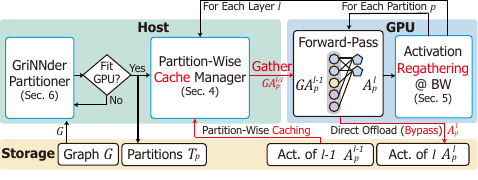}
    \vspace{-7mm}
    \caption{Overall workflow with cache-(re)gather-bypass.}
    \label{fig:overview}
    \vspace{-3mm}
\end{figure}

To address the limitations identified in \cref{sec:background}, we propose \thiswork, the first framework enabling storage-offloaded full-graph GNN training to break the GPU and host memory wall.
As \naive storage employment demonstrates, efficient storage-based full-graph training requires a fundamentally different approach than simple offloading.
We introduce \textit{\genstructure}, a systematic framework for orchestrating data movement across the GPU-host-storage hierarchy, specifically designed for full-graph GNN training.

The core of \genstructure is the \textbf{\textit{cache-(re)gather-bypass}} mechanism.
The workflow is illustrated in \cref{fig:overview} (see \cref{app:algo} for the procedure).
Assuming that the graph $G$ is partitioned into small subgraphs ($T_p$) (\cref{sec:partitioning}), the workflow is organized as follows:
\begin{itemize}[leftmargin=*, itemindent=.8em, itemsep=.5pt, topsep=.5pt]
\item \textbf{\textit{Cache.}} To avoid frequent fine-grained random accesses, the vertex activations are loaded from storage and \textbf{\emph{cache}}d in the host memory at per-partition and per-layer granularity.
For processing layer $l$, only the partitions from layer $l-1$ need to be accessed from the storage.
Thus, we can significantly reduce the working set, enabling efficient caching despite the limited capacity of host memory (\cref{sec:caching}).
\item \textbf{\textit{(Re)gather.}} In the GPU, the vertices in a single destination partition of layer $l$ are processed in a batch, which requires their source vertex activations in layer $l-1$.
Because transferring the source vertex data in a partition granularity would be too costly, the host processor \emph{\textbf{gather}}s and transfers the union of the source vertices of all destination vertices in the current partition ($GA_p^{l-1}$), from the cached data.
Unlike existing approaches, which snapshot the gathered input activation for the backward, we opt to \emph{\textbf{regather}} it just-in-time at the backward.
This significantly reduces the redundant I/O and memory pressure on the host/storage (\cref{sec:grad_engine}).
\item \textbf{\textit{Bypass.}} Meanwhile, some data do not benefit from caching, such as the topology and the resulting output activation ($A_p^l$) for the destination partition.
These data \emph{\textbf{bypass}} the host memory, and are written directly to the storage to prevent pressure on the host memory cache.
\end{itemize}

\begin{figure}[t]
    \centering
    \includegraphics[width=\columnwidth]{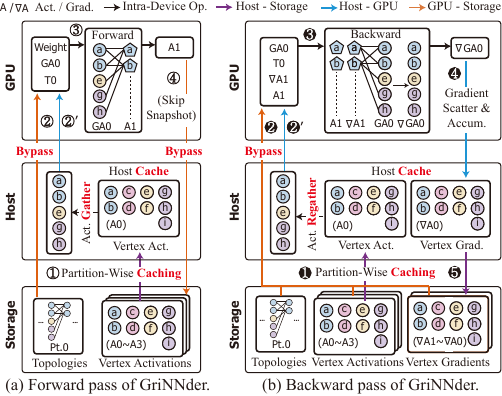}
    \vspace{-6mm}
    \caption{\thiswork forward and backward procedures for layer 1.
    }
    \vspace{-3mm}
    \label{fig:forward}
\end{figure}

\textbf{Detailed Procedure.}
\cref{fig:forward} details the forward/backward passes (we use $A0$ to denote $A^0$ for brevity).
\cref{fig:forward}a depicts the forward pass for partition $0$ of layer $1$.
\circled{1} (\textbf{\textit{Cache}}) Layer 0 activations ($A0$) are loaded into the host cache at partition granularity.
\circled{2} Partitioned graph structure $T_{0}$ uploads to GPU.
\circled{2}' (\textbf{\textit{Gather}}) Required vertex activations $GA0$ are gathered in host memory and transferred to GPU. 
\circled{3} GPU executes the forward pass to output $A1$.
\circled{4} (\textbf{\textit{Bypass}}) Computed activations $A1$ are offloaded to storage via GPU Direct Storage (GDS)~\cite{gds}, skipping snapshots to reduce redundancy (\cref{sec:grad_engine}).

\cref{fig:forward}b illustrates backward for the same partition.
The procedure mirrors forward in reverse with added complexity from activation gradients ($\nabla A1$, $\nabla GA0$).
\bcircled{1} (\textbf{\textit{Cache}}) Activations ($A0$) are cached in host memory partition-wise for frequent reuse.
\bcircled{2} (\textbf{\textit{Bypass}}) Activations $A1$ and gradients $\nabla A1$ load directly from storage.
Backward takes $A1$, $\nabla A1$, $GA0$ as inputs producing $\nabla GA0$.
\bcircled{2}' (\textbf{\textit{Regather}}) $GA0$ is fetched from the host cache through regathering, not snapshots (\cref{sec:grad_engine}).
\bcircled{3} GPU computes activation gradients ($\nabla GA0$) using loaded activations/gradients.
\bcircled{4} Source vertex gradients ($\nabla GA0$) update in host memory via scattered accumulation, ensuring correctness for vertices shared across partitions.
Host memory serves as a write-back buffer for vertex activation gradients.
\bcircled{5} After processing the entire layer, gradients are offloaded to storage.

\section{\OurStructure}
\label{sec:structure}
\label{sec:caching}

\textit{\textbf{Key takeaway}: 
Similar to vertex degrees in real-world graphs, cross-partition dependencies also follow a power-law distribution.
Exploiting this characteristic enables efficient storage I/O management through targeted caching.
}

\Genstructure manages host memory cache at \textit{partition} granularity, exploiting locality while avoiding fine-grained storage access patterns.
The cache replacement policy operates in two modes.
Under sufficient host memory capacity, the cache retains complete layers, maximizing reuse across partition iterations.
When host memory cannot accommodate all layers, the policy evicts entire layers in LRU order.
For extreme cases where a single layer exceeds host memory capacity (observed only in Papers dataset in \cref{sec:eval:main} and IGBM with reduced cache in \cref{sec:eval:swap}), the policy degrades gracefully to partition-wise eviction.
We now present the design rationale and mechanisms underlying this hierarchical cache management strategy.

\begin{figure}[t]
    \centering
    \includegraphics[width=\columnwidth]{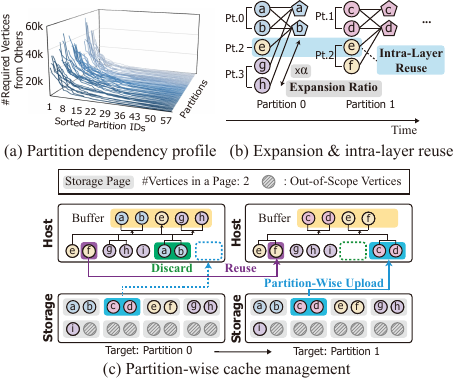}
    \vspace{-7.5mm}
    \caption{Details and rationales of \ourstructure.
    \texttt{Pt.1} denotes \texttt{Partition 1}.}
    \vspace{-4mm}
    \label{fig:cache}
\end{figure}

Cross-partition access patterns exhibit power-law distributions analogous to vertex degree distributions in real-world graphs, emerging from inherent clustering tendencies~\cite{leskovec}.
\cref{fig:cache}a validates this on the IGBM dataset.
For each partition (y-axis), we measure the required vertices from other partitions.
Sorting these counts (x-axis) reveals that among 64 partitions, dependencies concentrate within approximately 10 partitions (additional datasets in \cref{app:partition_profile}).
We exploit this skewed distribution through two mechanisms.

\textbf{Layer-Wise Partition Caching}: 
Within a layer, partitions share activations and gradients due to cross-partition edges.
In \cref{fig:cache}b, vertex $e$'s activation is used in both partitions $0$ and $1$.
With average expansion ratio $\alpha$ (\#required/\#target), activations are reused $\alpha-1$ times within that layer, causing redundant storage accesses.
To mitigate this inefficiency, we introduce \emph{intra-layer reuse}, caching frequently reused partitions in host memory.
Data with minimal intra-layer reuse (graph topology, output activations) are placed in storage and bypass host memory using GPUDirect Storage~\cite{gds} (GDS), reducing I/O traffic and cache conflicts.
Note that this design can be used in general even when GDS is unavailable (see \cref{app:gds}).

\textbf{Partition-Wise Cache Management}:
\thiswork uses partitions as the load/evict granularity for host memory cache. 
Alternatively, vertex granularity would require reading single vertex features (64--1,024B) on cache misses.
Since storage devices access data at page granularity (e.g., 16KiB), this incurs substantial unnecessary I/O.
Partition-wise management reduces this overhead as partition sizes are typically a few GBs.
For instance, processing partition 0 (vertices $a$, $b$) with dependencies to $a$, $b$, $e$, $g$, $h$ loads three partitions (0, 2, 3) to host memory (\cref{fig:cache}c).
For partition 1 with dependencies to $c$, $d$, $e$, $f$ spanning partitions 1 and 2, we reuse cached partition 2 and only load partition 1, evicting unused partition 0.
This reuses vertex features without fine-grained random accesses.
During this procedure, we keep a small buffer to send each partition's input activations from the host to the GPU.
In the worst case, partition-wise management incurs overhead when dependencies are uniformly distributed across many partitions.
However, as shown in \cref{fig:cache}a, the dependencies within a partition are concentrated in a few partitions, enabling stable caching performance.
For the detailed comparison between the partition-wise and vertex-wise management, see \cref{app:vertex_vs_partition}.
We further minimize latency by overlapping processing with cache management and maximizing sequential GPU access (\cref{app:io_opt}).

\section{\OurGrad}
\label{sec:grad_engine}
\label{sec:ioopt}

\textit{\textbf{Key takeaway}: 
PyTorch's autograd engine requires redundant snapshot storage, causing $\alpha$-fold\footnotemark[1] data amplification.
Our \ourgrad, a regathering-based gradient engine for just-in-time activation reconstruction, eliminates snapshots and reduces storage I/O.
}

One of the key challenges for employing storage in full-graph GNN training is \emph{data amplification}, where repeated snapshots of input activations inflate memory capacity and storage I/O demands.
As described in the previous section, \thiswork partitions the graph and caches graph features/gradients in the host memory at partition granularity.

\begin{figure}[t]
    \centering
    \includegraphics[width=\columnwidth]{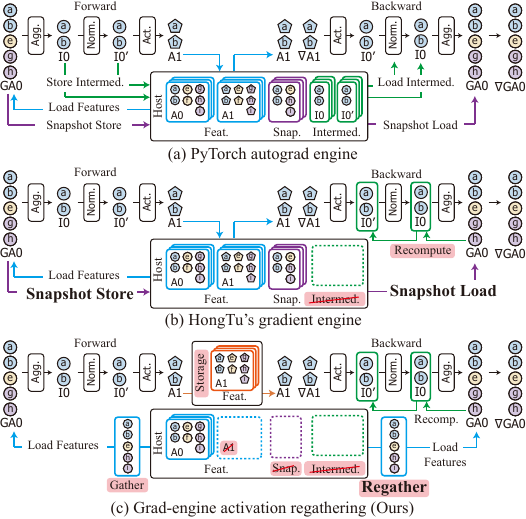}
    \vspace{-7mm}
\caption{Advantage of (c) \ourgrad compared to (a) PyTorch autograd and (b) HongTu.
    }
    \vspace{-5mm}
    \label{fig:our_grad}
\end{figure}

\footnotetext[1]{While $\alpha$, the average expansion ratio of partitions (\#required/\#target), was used to express the reusability in \cref{sec:caching}, it becomes a critical amplification factor here.}

However, existing autograd engines~\cite{pytorch, pyg} such as \texttt{torch.autograd} from PyTorch, even with generic activation checkpointing~\cite{gradient_checkpoint}, are not designed with such optimizations and require substantial host memory when employing offloading, as illustrated in \cref{fig:our_grad}a.
The vanilla autograd engine stores activation snapshots (`Snap.') and intermediate snapshots of all operations (`Intermed.'), such as normalization ($I0$) and activation function ($I0'$). 
While this design is reasonable for vision or language models with bounded activation sizes, it incurs significant memory overhead for GNNs, where activation (snapshot) sizes scale with graph size and the number of partitions with the magnitude of $\alpha$.

The amplification problem reveals three core challenges.
First, GNN propagation requires graph-topology-aware reconstruction rather than generic recomputation of tensor operations.
Second, partition-wise caching necessitates coordinated optimization across the GPU-host-storage memory hierarchy.
Third, the $\alpha$-fold snapshot amplification (where $\alpha \approx 8$ for large graphs) incurs a huge memory footprint, as all partition snapshots are required during backpropagation.
Prior techniques address related but distinct problems.
Generic activation checkpointing~\cite{gradient_checkpoint} trades computation for memory but cannot account for graph topology or partition granularity.
HongTu~\cite{hongtu} recomputes intermediate activations but still requires storing $\alpha$-fold amplified snapshots, which will be discussed in the next paragraph.
Ginex~\cite{ginex} optimizes cache management for mini-batch training but does not address full-graph training's snapshot amplification.

HongTu~\cite{hongtu} (\cref{fig:our_grad}b) mitigates PyTorch autograd's issue of snapshotting intermediates (e.g., $I0$ and $I0'$) by recomputing intermediate activations on demand.
It employs a gradient engine to snapshot the gathered activations to reduce latency (by enabling sequential access to snapshots), but at the cost of increased snapshots and redundant vertex data across partitions. 
As a result, each vertex may be stored up to $\alpha$ times, which adversely impacts memory consumption and bandwidth requirements, particularly for large datasets.
This is because it assumes abundant host memory and does not consider the use of storage, which has much lower bandwidth than host memory.

To address these limitations, we introduce \emph{\ourgrad}, illustrated in \cref{fig:our_grad}c, which eliminates snapshot redundancy through three optimizations.
First, we observe that the activation snapshot $GA0$ is essentially a reorganization of activations $A0$ according to graph topology.
Rather than pre-storing these reorganized snapshots, we \emph{regather} them on-demand during backpropagation from the original activations $A0$ by applying the gather operation.
While this introduces an additional regather operation at the host, it eliminates snapshot storage (proportional to $\alpha$) and the associated I/O overhead.
Critically, since storing all snapshots would overflow typical host memory capacity, pre-storing snapshots mandates additional storage I/O, which incurs far greater overhead than our regathering approach.
Second, intermediate values are removed from the host memory and recomputed just in time from the regathered $GA0$.
In the figure, $I0$ is recomputed by aggregation using the topology, and $I0'$ is obtained by further applying normalization. 
This shares the same principle as existing recomputation techniques~\cite{hongtu, gradient_checkpoint}, but is merged with the regathering mechanism.
Finally, to further reduce host memory pressure, the output feature $A1$ is bypassed and written directly to storage.
The combination of these optimizations enables \ourgrad to operate with minimal host memory footprint while substantially reducing storage I/O volume.

It is also worth mentioning HongTu's additional optimization, which is specific to GCN-like models.
On top of the default HongTu engine depicted in \cref{fig:our_grad}b, the optimization stores a snapshot of the aggregated intermediate activations ($I0$) instead of gathered activations ($GA0$) to reduce host-GPU traffic from $\alpha D$ to $D$.
This is only possible because $\nabla GA0$ can be directly computed from $I0$ in GCN.
Therefore, this is infeasible for GAT-like models that apply additional attention operations during aggregation.
Moreover, this additional method still requires $D \times L$ of additional host memory ($D=|V||H|$). In host-memory-constrained settings, this can easily cause OOM, thereby converting host-GPU traffic into slower storage traffic.

We can analyze the advantage of \ourgrad compared to this additional optimization as follows.
In the forward pass, HongTu is slower than \thiswork because it additionally snapshots $D$ activations, whereas \ourgrad eliminates snapshotting by regathering in the backward pass.
In the backward pass, assuming HongTu's $D \times L$ additional memory usage forces host-GPU traffic to spill into slow OS swap storage traffic, HongTu loads $2D$ activations and $D$ gradients from storage, while offloading $\alpha D$ gradients to storage.
On the other hand, \thiswork loads $(\alpha +1)D$ activations and $D$ gradients from host memory, offloading $\alpha D$ gradients to host memory.
Omitting computation, which is often negligible compared to I/O in resource-limited environments, and denoting host-GPU bandwidth as $B_{host}$ (typically x16 lanes) and SSD bandwidth as $B_{SSD}$ (typically x4 lanes), \thiswork is faster in the backward pass if:
\begin{equation}
\begin{aligned}
(2D+D+\alpha D)/B_{SSD}>[(\alpha+1)D+D+\alpha D]/B_{host} \notag
\end{aligned}
\end{equation}
This simplifies to $B_{host}/B_{SSD}>2(\alpha+1)/(\alpha+3)\approx1.2-1.6$ for practical $\alpha=2-8$.
Since physical lane differences typically yield $B_{host}/B_{SSD} \geq 2-4$, \ourgrad is usually preferable.
HongTu's intermediate snapshotting is only effective when host memory is abundant or when graphs are small (i.e., when the host memory can accommodate the additional $D \times L$).

\textbf{I/O Volume and Memory Footprint.} Let $D=|V||H|$.
During the forward pass of a layer, the baseline autograd engine consumes $(2\alpha+3)D$ traffic between the GPU and the host, for the gathered input activations ($\alpha D$), snapshots ($\alpha D$), intermediate values ($2D$), and outputs ($D$).
Since the baseline easily exceeds the host memory limit, it mandates using OS swap memory with storage, and most of that traffic becomes the IO between the GPU/host and the storage.
\thiswork only consumes $\alpha D$ between the GPU and the host, $D$ between the GPU and the storage, and $D$ between the host and the storage while caching (when only cold misses exist).
In other words, while the baseline suffers from huge and slow storage traffic proportional to $\alpha$, \ourgrad only requires a $2D$ amount of storage traffic.
In terms of the memory footprint during the forward, the baseline stores snapshots $(\alpha D)$, activations ($D$), and intermediate values ($2D$) per layer.
On the other hand, \ourgrad only occupies $D$ space in the host memory, and $D$ in the storage for the outputs without redundancy.
In the worst case, where the baseline utilizes OS swap memory with storage, this represents a $\frac{2\alpha+3}{2} \approx 8.5\times$ reduction in storage I/O for typical $\alpha \approx 8$.
For more in-depth analyses (with another baseline~\cite{hongtu}), see \cref{app:indepth_io}.

\section{\OurPart}
\label{sec:partitioning}

\textit{\textbf{Key takeaway}:
Existing partitioning algorithms (e.g., METIS-based) often incur a significant memory footprint, harming the practicality of iterative partitioning workflows in full-graph GNN training environments.
}

\begin{figure}[t]
    \centering
    \includegraphics[width=\columnwidth]{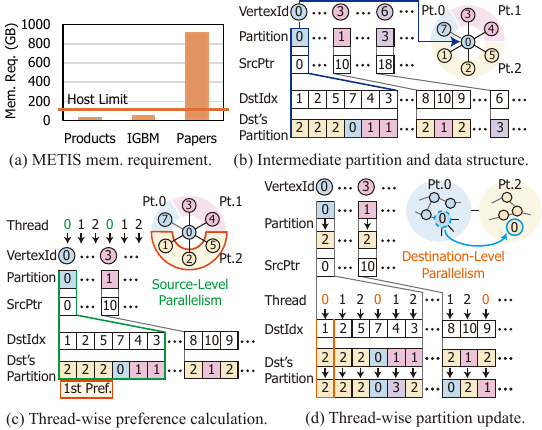}
    \vspace{-6mm}
    \caption{Motivation and a high-level overview of \ourpart.
    }
    \vspace{-4mm}
    \label{fig:partition}
\end{figure}

Graph partitioning is a critical enabler that allows \thiswork to efficiently utilize GPU memory, manage caches with minimal storage bandwidth, and minimize the traffic between host and GPU by reducing the expansion ratio ($\alpha$).
Although existing partitioners used in GNN domains produce near-optimal partitions, they often exceed single-server memory limits (\cref{fig:partition}a, measured with MT-METIS~\cite{mtmetis}) for large datasets such as Papers~\cite{ogb}.
Crucially, partitioning must be performed iteratively to find an adequate \#partitions that fit within GPU memory constraints---a process that becomes prohibitively expensive when each iteration requires external servers with sufficient host memory.
This could harm the practicality of full-graph GNN training workflows, clearly demonstrating the need for a lightweight partitioner.

Inspired by streaming partitioning approaches for distributed cloud systems (Spinner~\cite{spinner}), we devise a lightweight \ourpart with low memory consumption.
The key is to minimize the use of auxiliary data structures, whose size often largely surpasses that of the graph itself.
From an arbitrary initial partition, we iteratively refine the partition assignments to reduce the number of dependent partitions until convergence.
Detailed procedures and design insights are provided in \cref{app:part}.

\textbf{Algorithm Overview.}
\cref{fig:partition} outlines \ourpart's procedure and implementation, with exact preference scores and objective functions omitted for clarity.
At a high level, the algorithm attempts to move vertices to the partition with the most neighbors to reduce the number of dependent partitions while keeping partition sizes similar. 
\cref{fig:partition}b illustrates an example intermediate state during partitioning.
Following the CSR format, our data structure comprises source pointers (\texttt{SrcPtr}) and destination indices (\texttt{DstIdx}).
We manage an additional array (\texttt{Dst's Partition}) and fill this array with the partition ID of each destination index in \texttt{DstIdx}.
For example in \cref{fig:partition}b, vertex $0$ has neighbors $\{1,2,5,7,4,3\}$, and we fill \texttt{Dst's Partition} with their partition IDs: $\{2,2,2,0,1,1\}$.

In this state, vertex $0$ prefers partition $2$ (denoted as `1st Pref.' in \cref{fig:partition}c) because most of its neighbors reside in partition $2$.
We compute such preferences for each source vertex in parallel using source-level parallelism without additional memory usage, as illustrated in \cref{fig:partition}c.
After computing preferences, we relocate each source vertex to its preferred partition (label propagation).
For example, vertex $0$ moves to partition $2$, as depicted in \cref{fig:partition}d.
Since vertex $0$ now belongs to partition $2$, all entries in \texttt{Dst's Partition} pointing to vertex $0$ must be updated to reflect this change.
We perform this update efficiently using destination-level parallelism (\cref{fig:partition}d), where threads independently update entries corresponding to different destination vertices.
By iteratively updating preferred partitions following this procedure until convergence, we minimize the average expansion ratio ($\alpha$) across partitions.

For the initial state, we randomly assign vertices to partitions and observe stable behavior across runs, with partition quality largely insensitive to this random initialization.
In other words, while starting from a good initial state could reduce the number of iterations required for convergence, the final converged state is empirically independent of the initial state.
Additionally, the partition sizes across partitions are balanced via an explicit size-based penalty in the partition scoring function as follows (see \cref{app:part} for details).
Given state $S_{i}$ and partition $P_{j}$:
\begin{equation}
\begin{aligned}
Penalty_{(i,j)}=|P_{j}|/(\alpha_{balance}\times|V|/p),(0\leq j<p) \notag
\end{aligned}
\end{equation}
where $|P_{j}|$ is the current partition size, $|V|$ is the total number of vertices, $p$ is the number of partitions, and $\alpha_{balance}$ controls balance strictness (default $\alpha_{balance}=1.1$, allowing partitions to be $\sim$10\% larger than the equal size).
This discourages the growth of oversized partitions and maintains balance during partitioning.

\textbf{Memory Usage and Convergence.}
\Ourpart uses only a CSR representation (\texttt{SrcPtr}, \texttt{DstIdx}) and a \texttt{Dst's Partition} array to record each neighbor's current partition.
This totals $\mathcal{O}(2|V| + 2|E|)$ space compared to METIS' $\mathcal{O}(2|V|+|E|+\sum_{i=1}^{S}\big({|E_{i}|+|V_{i}|}\big))$ requirement~\cite{metis_memory}, where $S$ is the number of partitioning stages in METIS.
In practice, this achieves 7.10--24.37$\times$ memory reduction on large graphs (\cref{tab:part_mem_eval}).
\Ourpart converges in 30--50 iterations, consuming only 0.07\%/0.02\%/0.39\% of total training time on Products/IGBM/Papers datasets (\cref{app:part_converge})---a negligible overhead for practical workflows.
Despite its lightweight design, \ourpart achieves competitive partitioning quality compared to state-of-the-art lightweight partitioners (\cref{sec:eval:partitioning}).

\textbf{Integration with Training.}
We use METIS when host memory is sufficient, as it produces high-quality partitions.
However, when memory constraints mandate partitioning on external infrastructure, \ourpart offers a fast and memory-efficient alternative with competitive partition quality.
For detailed comparisons with Spinner~\cite{spinner} and state-of-the-art out-of-core partitioner (2PS-L~\cite{2ps_l}), see \cref{sec:eval:partitioning}.

\section{\thiswork API and Implementation}
\label{sec:framework}

We implement \thiswork as an extension to PyTorch's \texttt{torch.nn.Module}~\cite{pytorch}, providing a minimal-modification API for existing PyG~\cite{pyg} applications.
Users inherit the \texttt{GriNNderGNN} base class and implement a single \texttt{layer\_forward} method to enable layer-wise execution required for partition-based full-graph training.
This design decouples GNN model logic from the underlying offloading infrastructure, requiring typically several lines of code changes from standard PyG implementations (see \cref{app:api} for API details).

\begin{figure}[t]
    \centering
    \includegraphics[width=\columnwidth]{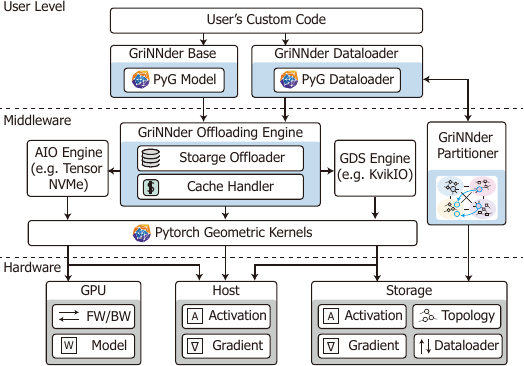}
    \vspace{-5mm}
    \caption{Framework structure of \thiswork.
    }
    \label{fig:framework}
    \vspace{-5mm}
\end{figure}

\cref{fig:framework} illustrates the framework architecture.
The \textbf{user-level} provides the base GNN module and a custom dataloader that manages partition-aware data serving, maintains cross-partition dependency metadata, and coordinates I/O scheduling.
The \textbf{middleware} integrates partition-wise caching (\cref{sec:structure}), gradient regathering (\cref{sec:grad_engine}), and lightweight partitioning (\cref{sec:partitioning}).
It implements the offloading engine, which orchestrates data movement through two specialized I/O engines: the Linux AIO interface wrapped by TensorNVMe~\cite{tensornvme} for host-storage transfers, and Kvikio~\cite{kvikio} for GPU-storage transfers via GPUDirect Storage (GDS)~\cite{gds}.

The engine tracks each activation's location and coordinates I/O operations across three \textbf{hardware} tiers.
Since offloaded training is typically I/O-bound, we implement aggressive I/O overlap to hide transfer latency.
Leveraging bidirectional PCIe bandwidth, the offloading engine pipelines the write of the previous partition's activations with the prefetch of the current partition's required activations.
We implement the dataloader and partitioner in C++ for high performance and expose them to Python through pybind11~\cite{pybind11}.

\section{Evaluation}
\label{sec:exp} 

\subsection{Experimental Settings and Baselines}
\label{sec:env}

\textbf{Hardware}: We run main experiments on a single GPU workstation with an AMD Ryzen9 7950X 3D CPU (16C 32T), 128GB DDR5-5600 RAM, one RTX A5000 (24GB) GPU, a PCIe 5.0 NVMe SSD (4TB), and a total 4TB swap space for swap-based experiments.
We chose a single GPU setup to demonstrate how \thiswork breaks through the host/GPU memory limitations.
For the multi-GPU extension, we utilized a multi-GPU server with four RTX4090 GPUs, 2$\times$Intel Xeon Gold 6442Y, 512GB DDR5 DRAM, and 2TB PCIe5.0 NVMe SSD.
For distributed baselines, we use a 4-server cluster; each node has four RTX A6000 GPUs interconnected by NVLink~\cite{nvbridge} and Infiniband SDR~\cite{infiniband}.
For IGBM/Papers, we needed all 16 GPUs to fit the data in the GPU memory.
For Products, using fewer GPUs could yield better performance, but we used all GPUs to maintain consistency among datasets.

\textbf{Models/Datasets}: We use 3-/5-layer GCN~\cite{gcn} with a hidden dimension of 256.
We also test GAT~\cite{gat} and GraphSAGE~\cite{graphsage}.
Datasets range from medium to large scale: Products~\cite{ogb}, IGBM~\cite{igb}, and Papers~\cite{ogb}.
We also utilized Kronecker graphs~\cite{kronecker} (average degree=10) with random initial features of dimension 128 and \#classes of ten.

\textbf{Baselines}: \textit{\textbf{(Training)}}
We compare \thiswork (GRD) against various single-server/distributed methods (detailed in \cref{app:prev_works}):
\circled{1} Micro-batch training (Betty \cite{betty}),
\circled{2} Micro-batch training with storage extension (Ginex \cite{ginex}),
\circled{3} Host memory offloaded training (HongTu \cite{hongtu}) with OS swap memory,
\circled{4} Distributed full-graph training (CAGNET \cite{cagnet}),
\circled{5} Distributed full-graph training with communication skipping (Sancus \cite{sancus}),
\circled{6} \Naive storage extension of full-graph training (ROC \cite{roc}).
We showed \circled{6} only in \cref{app:naive_storage} due to its much slower performance.
In the appendix, we also tested two storage-based mini-batch training (\circled{7} DiskGNN~\cite{diskgnn}, \circled{8} GNNDrive~\cite{gnndrive}) with micro-batch extension\footnotemark[2] (\cref{app:extension_limit}).
For out-of-memory issues in distributed baselines, we add host-memory checkpointing ($^*$) to enable execution.
Since \thiswork does not change the training algorithm itself, \thiswork achieves equal final accuracy with all the baselines (see \cref{sec:accuracy}) except \circled{5}, which is non-exact due to its staleness.
All baselines use the state-of-the-art partitioner MT-METIS~\cite{mtmetis}.
For fairness, if MT-METIS exceeds our setting's memory, we assume it was preprocessed elsewhere following standard practice, except for partitioning experiments.
\textit{\textbf{(Partitioning)}}
We also compared \ourpart with other alternative lightweight partitioners.
We chose Spinner~\cite{spinner} and 2PS-L~\cite{2ps_l}, a state-of-the-art out-of-core partitioner.

\footnotetext[2]{Since these systems are not designed for full-graph training workflows, extensions may deviate from their original performance characteristics.}

\subsection{Large Graph Training Results}
\label{sec:eval:main}

\begin{table}[t]
\centering
\caption{\textbf{Results of training time (min)/epoch.}}
  \label{tab:speedup}
  \vspace{0.1in}
\setlength{\tabcolsep}{1pt}
\resizebox{\columnwidth}{!}
{
    \begin{tabular}{lllccc}
      \toprule
      & & \textbf{\#\,nodes} & \footnotesize{2.4M} & \footnotesize{10M} & \footnotesize{100M} \\
      & & \textbf{Method} & ~~~~~~\textsc{Products}~~~~~~ & ~~~~~~\textsc{IGBM}~~~~~~ & ~~~~~~\textsc{Papers}~~~~~~ \\
      \midrule
      \mr{6}{\rotatebox{90}{\small{$|L|=3$}}} & ~\mr{4}{\rotatebox{90}{\small{Limited}}}
       & ~~\textsc{Betty}    ~& 0.61 & 28.71 & \small{GPU OOM} \\
       & & ~~\textsc{Ginex}  ~& 9.00 & \small{GPU OOM} & 17.72 \\
       & & ~~\textsc{HongTu} ~& 0.17 & 6.46 & \small{Swap OOM} \\
      & & \cellcolor{gray!15}~~\textbf{GRD}     ~& \cellcolor{gray!15}\textbf{0.12} & \cellcolor{gray!15}\textbf{0.93} & \cellcolor{gray!15}\textbf{9.07} \\
      \cmidrule(lr){2-6}
      & ~\mr{2}{\rotatebox{90}{\small{Dist.}}}
      & ~~\textsc{Cagnet}    ~& 0.21 & 1.41 & $^{*}$10.01 \\
      & & ~~\textsc{Sancus}  ~& 0.19 & $^{*}$0.77 & \small{$^*$GPU OOM} \\
      \cmidrule(lr){1-6}
      \mr{6}{\rotatebox{90}{\small{$|L|=5$}}} & ~\mr{4}{\rotatebox{90}{\small{Limited}}}
       & ~~\textsc{Betty}   ~& 1.05 & \small{GPU OOM} & \small{GPU OOM} \\
       & & ~~\textsc{Ginex} ~& 15.10 & \small{GPU OOM} & \small{GPU OOM} \\
       & & ~~\textsc{HongTu}~& 0.32 & 14.90 & \small{Swap OOM} \\
       &  & \cellcolor{gray!15}~~\textbf{GRD} ~& \cellcolor{gray!15}\textbf{0.23} & \cellcolor{gray!15}\textbf{1.52} & \cellcolor{gray!15}\textbf{12.03} \\
      \cmidrule(lr){2-6}
      & ~\mr{2}{\rotatebox{90}{\small{Dist.}}}
      & ~~\textsc{Cagnet}  ~& 0.38 & 2.10 & \small{$^*$GPU OOM} \\
      & & ~~\textsc{Sancus} ~& 0.36 & $^*$1.41 & \small{$^*$GPU OOM} \\
      \bottomrule
      \multicolumn{6}{r}{\textsc{Sancus}: Non-exact full-graph (with staleness)}
    \end{tabular}
}

\end{table}

\begin{table}[t]
\centering
\vspace{-5mm}
\caption{\textbf{Training time (min)/epoch sensitivity for graph sizes with synthetic graphs.}
For results with ablation, see \cref{app:synth}.}
  \label{tab:synth_speedup}
  \vspace{0.1in}
\setlength{\tabcolsep}{1.5pt}
  \resizebox{\columnwidth}{!}{%
    \begin{tabular}{llcccc}
      \toprule
      & \textbf{\#\,nodes} & ~~~~~\textsc{4.2M}~~~~~ & ~~~~~\textsc{8.4M}~~~~~ & ~~~~~\textsc{16.8M}~~~~~ & ~~~~~\textsc{33.6M}~~~~~ \\
      \midrule
      \mr{2}{\rotatebox{90}{\small{$|L|$=3}}}
       & ~~\textsc{HongTu} ~& 0.43 & 0.83 & 7.25 & 36.31 \\
       & \cellcolor{gray!15}~~\textbf{GRD}     ~& \cellcolor{gray!15}\textbf{0.29} & \cellcolor{gray!15}\textbf{0.59} & \cellcolor{gray!15}\textbf{1.93} & \cellcolor{gray!15}\textbf{3.73} \\
      \cmidrule(lr){1-6}
      \mr{2}{\rotatebox{90}{\small{$|L|$=5}}}
       & ~~\textsc{HongTu}~& 0.83 & 1.99 & 19.15 & 96.99 \\
       & \cellcolor{gray!15}~~\textbf{GRD} ~& \cellcolor{gray!15}\textbf{0.57} & \cellcolor{gray!15}\textbf{1.14} & \cellcolor{gray!15}\textbf{3.71} & \cellcolor{gray!15}\textbf{7.76} \\
      \bottomrule
    \end{tabular}
  }

\end{table}

\cref{tab:speedup} presents per-epoch training time for \thiswork (GRD) compared to five baselines—Betty, Ginex, HongTu, CAGNET, and Sancus—using 3-/5-layer GCNs (hidden dimension 256) on Products, IGBM, and Papers.

\textbf{Micro-Batch (Betty, Ginex)}: Despite Betty’s memory-only design (no storage), GRD achieves up to 30.98$\times$ faster training, largely due to Betty’s redundant computation from the message flow graph’s neighbor explosion.
Ginex uses storage to extract message flow graphs, yet still suffers from the same issue, which GRD improves by up to 77.92$\times$.

\textbf{Products (Medium)}: Since HongTu can fit Products entirely in host memory, one might expect it to outperform storage-based GRD.
In practice, HongTu’s redundant snapshots slow it down, allowing GRD to beat it by 1.44/1.40$\times$ on 3-/5-layer GCNs.

\textbf{IGBM (Large)}: Micro-batch methods suffer from GPU OOM on deeper models—Betty/Ginex often cannot handle the neighbor explosion.
HongTu must manage large volumes of data in host memory, drastically increasing overhead.
In contrast, GRD is 6.97/9.78$\times$ faster than HongTu with caching and non-redundancy.
Even against multi-GPU CAGNET, GRD achieves 1.52/1.38$\times$ speedup because the distributed baselines are bottlenecked by inter-server communication over a slow 10Gbps interconnect.

\textbf{Papers (100M)}: This highlights GRD’s scalability on larger datasets.
Betty and Ginex often fail on deeper models with OOM from neighbor explosion, and HongTu fails from activation snapshots. 
GRD avoids these with efficient caching and no redundant snapshots. 
Ginex can run the 3-layer model but is 1.95$\times$ slower than GRD.
Notably, GRD is faster than CAGNET (1.10$\times$) despite using a single GPU.

\textbf{Synthetic Graphs}: In \cref{tab:synth_speedup}, we tested various-sized Kronecker graphs to validate scalability, where \thiswork provides stable speedup over HongTu (1.41--12.50$\times$).

\begin{table}[t]
\centering
\caption{\textbf{Sensitivity on effective cache size with ablation (training time (min)/epoch).}
    }
  \label{tab:swap_hidden_sensi}
  \vspace{0.1in}
\setlength{\tabcolsep}{7pt}
  \resizebox{\columnwidth}{!}{%
    \begin{tabular}{llccc}
      \toprule
      & \footnotesize{\textbf{\# hiddens}} & \footnotesize{$|H|$=384} & \footnotesize{$|H|$=512} & \footnotesize{$|H|$=1024}  \\[-0.05cm]
      & \textbf{Method} & ~~~\textsc{0.75 \$ Size} & ~~~\textsc{0.5 \$ Size} & ~~~\textsc{0.25 \$ Size}  \\
      \midrule
      \mr{3}{\rotatebox{90}{\small{$|L|$=3}}}
       & ~~\textsc{HongTu}    ~& 12.53 & 18.67  & 39.32   \\
       & \cellcolor{gray!15}~~\textsc{GRD-G}  ~& \cellcolor{gray!15}\textbf{1.20} & \cellcolor{gray!15}\textbf{1.51}  & \cellcolor{gray!15}20.68 \\
       & \cellcolor{gray!15}~~\textsc{GRD-GC} ~& \cellcolor{gray!15}1.41 & \cellcolor{gray!15}1.91  & \cellcolor{gray!15}\textbf{3.98} \\
      \cmidrule(lr){1-5}
        \mr{3}{\rotatebox{90}{\small{$|L|$=5}}}
       & ~~\textsc{HongTu}    ~& 25.07 & 31.81  & 93.42  \\
       & \cellcolor{gray!15}~~\textsc{GRD-G}  ~& \cellcolor{gray!15}10.26 & \cellcolor{gray!15}12.50  & \cellcolor{gray!15}42.14 \\
       & \cellcolor{gray!15}~~\textsc{GRD-GC} ~& \cellcolor{gray!15}\textbf{2.54} & \cellcolor{gray!15}\textbf{3.37}  & \cellcolor{gray!15}\textbf{13.65} \\
      \bottomrule
    \end{tabular}
  }

\end{table}

\subsection{Ablation by Decreasing Effective Cache Size and Cache Hit Rate}
\label{sec:eval:swap}

\cref{tab:swap_hidden_sensi} analyzes \thiswork’s sensitivity to effective cache size by varying the hidden dimension on IGBM.
We ablated \thiswork: HongTu, HongTu + \ourgrad (GRD-G), and GRD-G + \ourstructure (GRD-GC).
\thiswork outperforms HongTu by 6.84–12.34$\times$.
When host memory can cache most data (in 3 layers), GRD-G alone provides sufficient performance benefits.
However, in 5-layer settings, host memory becomes a bottleneck, making cache replacement crucial.
Thus, GRD-GC gains 3.09--4.04$\times$ speedup over GRD-G.
Overall, \thiswork is robust on cache sizes.
Also, we find that larger datasets have higher cache hit rates (53.70--92.77\%) with more reuse from the higher \#partitions in \cref{app:cache_hit}.

\subsection{Analysis on Host Memory Usage}
\begin{figure}[t]
    \centering
    \includegraphics[width=\columnwidth]{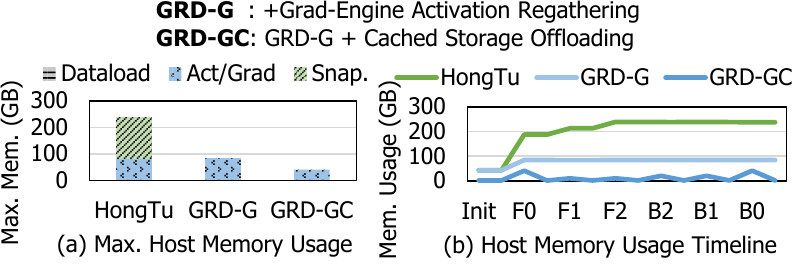}
    \vspace{-7mm}
    \caption{Host memory usage of \thiswork on the IGBM.}
    \label{fig:memory-grinnder}
    \vspace{-1mm}
\end{figure}

\cref{fig:memory-grinnder}a shows an ablation study on how \thiswork reduces host memory consumption.
We compare GRD-G (i.e., HongTu + \ourgrad) and GRD-GC (GRD-G + \ourstructure).
For ablation purposes, GRD-GC imposes an explicit cache cap of one layer's activations and gradients to demonstrate layer-wise caching on top of GRD-G.
HongTu suffers from snapshots, while GRD-G eliminates them.
GRD-GC's layer-wise up/offload further cuts the peak usage from HongTu by 5.75$\times$.
\cref{fig:memory-grinnder}b shows the host memory usage timeline.
GRD-GC shows significantly lower memory usage over all timelines.

\subsection{Analysis on Partitioning Algorithms}
\label{sec:eval:partitioning}

\begin{figure}
    \centering
    \includegraphics[width=\columnwidth]{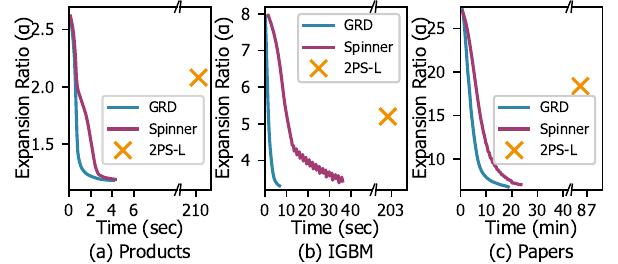}
    \vspace{-7mm}
    \caption{Time-to-quality comparison with alternatives on Products (4 parts), IGBM (32 parts), and Papers (2048 parts).}
    \label{fig:part_alternatives}
    \vspace{-3mm}
\end{figure}

\begin{figure}
    \centering
    \includegraphics[width=.95\columnwidth]{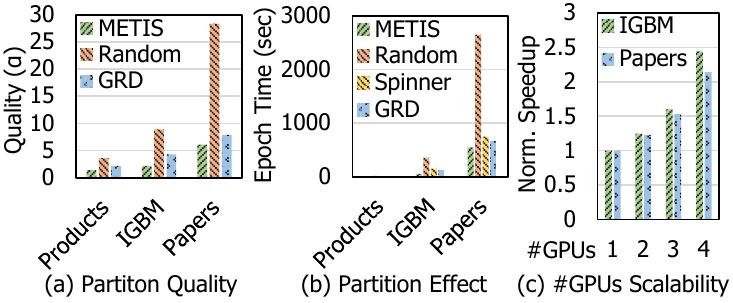}
    \vspace{-4mm}
    \caption{Partitioning quality comparison, effect of partitioning on training time, and multi-GPU scalability.
    }
    \label{fig:sensi}
    \vspace{-2mm}
\end{figure}

\textbf{Comparison with Alternatives and METIS}: We compared the time-to-quality (i.e., expansion ratio $\alpha$, lower is better) of \ourpart (GRD) with the famous streaming algorithm (Spinner) and state-of-the-art out-of-core partitioner (2PS-L~\cite{2ps_l}) in \cref{fig:part_alternatives}.
We ran 50 iterations for GRD/Spinner and used the default setting for 2PS-L.
GriNNder quickly and stably results in higher-quality partitions compared to the lightweight alternatives.
Furthermore, we compared the partition quality of \ourpart with METIS in \cref{fig:sensi}a for the same setup as \cref{fig:part_alternatives}.
While METIS achieves modestly better quality, it requires prohibitive memory usage -- often exceeding available host memory for large graphs.
Therefore, we use switching-aware partitioning when host memory is limited, as mentioned in \cref{sec:partitioning}.
\textbf{Convergence and Practical Overhead}:
We also reported the trend of partitioning quality improvement (convergence) and practical overhead of \ourpart in \cref{app:part_converge}.
We observed that at most 50 iterations are enough for convergence.
Also, the practical overhead was only 0.07/0.02/0.39\% of the total training time on Products/IGBM/Papers, respectively.

\textbf{Partitioning and Training Time}: 
Among datasets, only Papers exceeded the host memory capacity.
Partitioning it into 16 parts with MT-METIS triggers host swap due to its large memory demand and took 77.26 min, making \ourpart 10.51$\times$ faster (7.35 min).
We used \#partitions=16 only in this experiment because MT-METIS showed out-of-time (over three hours) for the default \#partitions=2048.
\cref{fig:sensi}b evaluates how partitions affect the training of 3-layer GCNs on Products/IGBM.
While MT-METIS with near-optimal partitions yields the shortest training time, it uses significantly more memory.
GRD needs far less memory while improving training speed by 1.59$\times$/2.80$\times$ on Products/IGBM over random partitioning.
Compared to Spinner, GRD provides better partitioning quality, thereby achieving up to a 1.20$\times$ training speedup.

\begin{table}[t]
    \centering
    \vspace{-3mm}
    \caption{\textbf{Memory usage (GB) of partitioning.}}
    \vspace{0.1in}
    \label{tab:part_mem_eval}
    \setlength{\tabcolsep}{7pt}
    \resizebox{\linewidth}{!}{%
      \begin{tabular}{llcccc}
        \toprule
        \textbf{Dataset} & \textbf{Method} & \textbf{Graph} & \textbf{Part.\ Label} & \textbf{Add.} & \textbf{Total}\\
        \midrule
        \multirow{2}{*}{\textsc{Products}} & MT-METIS & 1.01 & 0.01 & 9.93 & 10.95\\
                                           & \cellcolor{gray!15}\textbf{GRD}      & \cellcolor{gray!15}1.01 & \cellcolor{gray!15}0.01 & \cellcolor{gray!15}\textbf{0.52} & \cellcolor{gray!15}\textbf{1.54} \\
        \cmidrule(lr){1-6}
        \multirow{2}{*}{\textsc{IGBM}} & MT-METIS & 1.12 & 0.04 & 28.34 & 29.50 \\
                                       & \cellcolor{gray!15}\textbf{GRD}      & \cellcolor{gray!15}1.12 & \cellcolor{gray!15}0.04 & \cellcolor{gray!15}\textbf{0.87}  & \cellcolor{gray!15}\textbf{2.03} \\
        \cmidrule(lr){1-6}
        \multirow{2}{*}{\textsc{Papers}} & MT-METIS & 26.71 & 0.44 & 867.84 & 895.00 \\
                                         & \cellcolor{gray!15}\textbf{GRD}      & \cellcolor{gray!15}26.71 & \cellcolor{gray!15}0.44 & \cellcolor{gray!15}\textbf{9.56} & \cellcolor{gray!15}\textbf{36.72} \\
        \bottomrule
      \end{tabular}
      }
\end{table}

\textbf{Memory Usage}: \cref{tab:part_mem_eval} shows that \thiswork’s partitioning greatly reduces memory usage by 7.10–24.37$\times$ compared to MT-METIS.
MT-METIS requires additional memory for partitioning-stage-wise intermediates.
In contrast, \ourpart only needs $\mathcal{O}(|E|)$ extra space.
We excluded Spinner from this comparison because we ported it from its cloud-based implementation to a single-server setup, applying \thiswork's memory-optimized parallelization in our evaluations to ensure a fair comparison.

\subsection{Multi-GPU Scalability}
\label{sec:multi_gpu}

While evaluations were mainly conducted on a single GPU environment to demonstrate how \thiswork breaks through the host/GPU memory limitations, \thiswork is extendable to multi-GPU environments.
As discussed in \cref{app:multi_gpu}, GriNNder supports multi-GPU execution via partition parallelism, where partitions are divided into disjoint sets, and each GPU processes its assigned set independently.
Gradient synchronization is performed via CPU-side atomic vertex gradient accumulation and an all-reduce of weight gradients among GPUs before weight updates.
We tested the multi-GPU scalability of \thiswork in \cref{fig:sensi}c.
We excluded Products because it is too lightweight to be run on multiple GPUs.
The speedup is scalable to the number of GPUs, but some overhead is incurred due to the system's shared resources---host memory and storage bandwidth.

\subsection{Sensitivity to Model, Number of Layers, and Partition Configuration}
\begin{figure}
    \centering
    \includegraphics[width=.9\columnwidth]{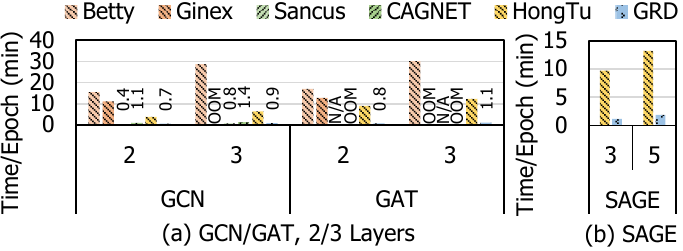}
    \vspace{-4mm}
    \caption{Sensitivity to model types and \#layers.
    Sancus does not provide GAT staleness implementation (denoted by N/A).
    }
    \vspace{-6mm}
    \label{fig:gat_layer_sensi}
\end{figure}

\cref{fig:gat_layer_sensi} shows the comparison on other models (GAT~\cite{gat} and GraphSAGE~\cite{graphsage}) and other numbers of layers, using IGBM.
\thiswork maintains consistent and significant speedups over baselines, demonstrating its efficiency beyond GCN and across varying numbers of layers.
\thiswork can also be extended for heterogeneous GNNs, showing stable performance (\cref{app:hetero}).
We examine the impact of \#partitions configuration in \cref{app:config_sensi}.
Compared to HongTu, from the efficient caching and redundancy elimination, \thiswork is much more robust to the \#partitions.

\subsection{Overhead of Regathering and Recomputation}
\begin{figure}
    \centering
    \includegraphics[width=.95\columnwidth]{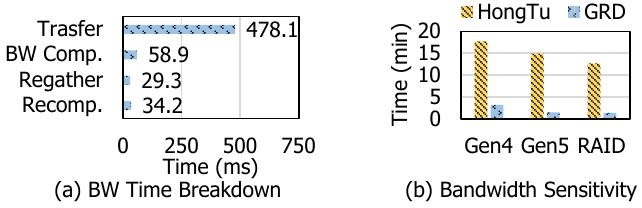}
    \vspace{-3mm}
    \caption{Profile of regather/recompute in the backward pass and SSD bandwidth sensitivity on training time/epoch.
    }
    \vspace{-5mm}
    \label{fig:overhead_bw_sensi}
\end{figure}

We quantify recomputation/regathering overheads in \cref{fig:overhead_bw_sensi}a, showing the time breakdown of a single backward pass for one partition at an intermediate layer in a 3-layer GCN on IGBM. Since host-GPU transfer dominates, regather accounts for 4.88\%, and recompute 5.69\% of the backward pass. Importantly, while measurable at the per-partition level, their impact on end-to-end training time is largely hidden by I/O overlap, as shown in \cref{fig:overlap_overview}.

\subsection{Storage Bandwidth and Endurance Analysis}
\textbf{Sensitivity to Storage Bandwidth.}
We evaluate SSD bandwidth sensitivity in \cref{fig:overhead_bw_sensi}b with PCIe Gen. 4 SSD ($\sim$7 GB/s read/write), PCIe Gen. 5 SSD (original, $\sim$12 GB/s read/write), and RAID5 (8× Intel D7-P5520 SSDs, $\sim$56.8 GB/s read, $\sim$25.9 GB/s write).
With lower-bandwidth SSDs, \thiswork remains robust, achieving significant speedup over HongTu. With near-DRAM bandwidth, speedup is not proportional to the bandwidth increase because \thiswork's caching shifts the bottleneck to host-GPU communication.

\textbf{SSD Write Volume and Endurance.}
We quantify SSD writes for \thiswork and HongTu using a 3-layer GCN trained on IGBM/Papers with the actual write profiles measured by \texttt{smartmontools}~\cite{smartmontools}.
HongTu writes 192.4GB/2.35TB and \thiswork writes 2.1GB/647.2GB per epoch on IGBM/Papers, respectively.
\thiswork incurs significantly fewer writes than HongTu on IGBM through hierarchical caching that prioritizes host memory and redundancy elimination.
On Papers, \thiswork shows higher writes due to limited host memory relative to graph size, but given that we can achieve full accuracy with 100 epochs (\cref{sec:accuracy}), a total of 64.72TB of writes is required -- only 0.23\% of the endurance of a single Intel D7-P5520 SSD ($\sim$28PBW) and 0.033\% for our 8-SSD RAID5 setup ($\sim$196PBW).
Write volume can be further reduced via staleness techniques~\cite{sancus} (e.g., 2$\times$ reduction with single-step staleness), gradient compression techniques~\cite{powersgd, optimus_cc}, or sparsity-based compression~\cite{sgcn}.

\section{Discussion}
\subsection{Distributed Baselines on Modern Hardware}
In main evaluations, the distributed baselines are primarily network-bottlenecked due to their broadcast pattern, and 10Gbps interconnect.
As a further discussion, we analytically project distributed baselines for modern hardware as follows.
We estimate CAGNET's performance with InfiniBand HDR (200Gbps) on IGBM 3-layer, where CAGNET and \thiswork achieve 1.41 and 0.93 minutes/epoch (\cref{tab:speedup}).
Profiling on CAGNET shows that 80\% of the execution time is from communication and 20\% is from computation and static overheads.
With 20$\times$ bandwidth improvement (10Gbps $\xrightarrow[]{}$ 200Gbps), the execution time of CAGNET can be projected as 0.338 minutes.
Therefore, the projected speedup of CAGNET over \thiswork is 2.75$\times$.
This assumes perfect bandwidth scaling and zero latency overhead -- an optimistic upper bound favoring the distributed baseline.
In the actual execution of GNN training, the broadcast operation is also affected by intra-server communication and would not scale perfectly.
In terms of costs, a four-server 16-GPU A6000 cluster with InfiniBand HDR costs $\sim$\$132K, while our single A5000 workstation costs $\sim$\$3.3K (40$\times$ less expensive).
While modern hardware can yield faster training as mentioned, \thiswork provides a cost-effective solution for resource-constrained settings where high-performance clusters are unavailable or prohibitively expensive.

\subsection{Host Memory Caching of Hot Vertices}
Host memory caching of hot vertices could reduce I/O by keeping frequently accessed activations in host memory.
However, \thiswork's caching operates at the partition level because hot-vertex caching requires vertex-level management, which incurs fine-grained storage access and read amplification due to storage-page granularity (see \cref{app:vertex_vs_partition}).
For power-law graphs, these overheads are expected to outweigh the benefits: although hot vertices are few, they are scattered across partitions, requiring fine-grained accesses.
Nevertheless, carefully combining partition-level caching with hot-vertex awareness is a promising future direction.

\subsection{Motivation for Single-GPU Training}
Unlike transformer training, which benefits from arithmetic-intensive operators that parallelize efficiently across GPUs, full-graph GNN training is dominated by highly irregular, fine-grained graph memory accesses.
Under such conditions, adding GPUs introduces high synchronization and communication costs, resulting in low scaling efficiency.
Although multi-GPU settings could be faster on high-speed interconnects, such inefficiency makes single-GPU solutions appealing for GNN training.
Furthermore, single-GPU approaches for LLMs remain actively studied~\cite{flexgen, zero_infinity, qlora, hilos} as a low-cost solution.

\section{Related Work}
\label{sec:related_work}

\noindent\textbf{Full-Graph GNN Training.}
Full-graph training processes entire graphs without sampling, preserving complete input information~\cite{pipegcn, roc, bns_gcn, neugraph}.
Distributed systems~\cite{sancus, cagnet, bgl, neutronstar} incur substantial communication overhead and require expensive multi-GPU clusters.
Single-server methods~\cite{betty, hongtu} remain constrained by GPU/host memory capacity.
Hardware acceleration~\cite{gnnear, holisticgnn, barad_dur} requires specialized devices.
\thiswork breaks the memory capacity wall using commodity systems with NVMe SSDs.

\noindent\textbf{Storage-Based GNN Training.}
Prior systems~\cite{ginex, diskgnn, gnndrive, marius, helios} target subgraph-based training, managing initial features on storage for sampled subgraphs that fit in GPU memory.
Full-graph training presents fundamentally different requirements, and \thiswork addresses them through caching and regathering-based gradient engine.

\noindent\textbf{Activation Management.}
Checkpointing (i.e., snapshotting)~\cite{gradient_checkpoint, checkfreq, justintime_checkpointing} trades computation with memory.
LLM works~\cite{zero_infinity, flexgen} extend this through activation offloading.
However, LLM activations exhibit sequential layer dependencies enabling straightforward layer-by-layer management, while GNNs exhibit graph-structured dependencies requiring gathering from multiple partitions based on topology.
Prior GNN checkpointing~\cite{understanding_gnn_checkpointing, hongtu, redundancy_free_gnn} targets in-memory scenarios, introduces massive storage redundancy, or addresses only computational redundancy.
\thiswork introduces a regathering mechanism to reduce storage I/O.

\noindent\textbf{Graph Partitioning.}
METIS~\cite{metis}, adopted in various works~\cite{pipegcn, sancus, bgl}, requires memory up to 13.8$\times$ the input graph size~\cite{metis_memory}, precluding its use in memory-limited settings.
Alternatives~\cite{neugraph, streaming_metis, fennel} assume sufficient memory (e.g., cloud environments) or produce lower-quality partitions.
\thiswork's lightweight partitioning operates with only a small working set of memory.


\section{Conclusion}
To our knowledge, \thiswork is the first to break the GPU and host memory capacity wall of full-graph GNN training with storage offloading.
\thiswork introduces \genstructure, a general framework for managing the GPU-host-storage memory hierarchy through coordinated cache, regather, and bypass mechanisms.
Its co-designed optimizations based on \genstructure enable up to 9.78$\times$ speedup over state-of-the-art baselines and throughput comparable to distributed systems.

\section*{Acknowledgements}

This work was primarily supported by 
Samsung Electronics. 
Jinho Lee is also funded by the National Research Foundation of Korea (NRF) grant funded by the Korea government (MSIT) (RS-2026-25495605), and
Institute of Information \& communications Technology Planning \& Evaluation (IITP)
(RS-2024-00395134, 
RS-2024-00347394). 

\bibliography{references}
\bibliographystyle{others/mlsys2025}

\appendix

\clearpage

\crefalias{section}{appendix}

\section{Survey of Conferences' Submission on GNN Domains}
\label{app:survey}

In our survey of NeurIPS/ICML/ICLR 2024 papers, a total of 76 papers are related to GNN domains.
In 76 papers, 44.7\% (34 papers) used full-graph training, and among them, 38.2\% (13 papers) directly reported out-of-memory.
In terms of experimental environments, a total of 62 papers reported their GPU environments, and 45 papers utilized a single GPU (72.6\%).
Also, some papers with full-graph training directly stated that larger-sized datasets can incur out-of-memory when running their experiments.
This shows the importance of enabling full-graph training of large graphs under limited resources (e.g., a single GPU).

\section{Limitations of Existing Full-Graph Training Methods}
\label{app:prev_works}

Full-graph GNN training processes all vertices’ activations and gradients in a single pass, requiring substantial memory.

\textbf{Distributed Training:} Distributed full-graph training~\cite{cagnet, sancus, roc, pipegcn, bns_gcn} scales the number of GPUs to meet the memory requirement.
However, it requires a costly multi-server cluster.
Moreover, it significantly suffers from communication bottlenecks, especially inter-server communication.
We break down the training time of 3-/5-layer GCN with the widely used distributed full-graph training methods (CAGNET~\cite{cagnet} and Sancus~\cite{sancus}) in the four-server setup utilized in the evaluation (\cref{sec:env}).
They both suffer from severe communication bottlenecks from 80\% to 98\% of the execution time.

\begin{figure}[t]
    \centering
    \includegraphics[width=\columnwidth]{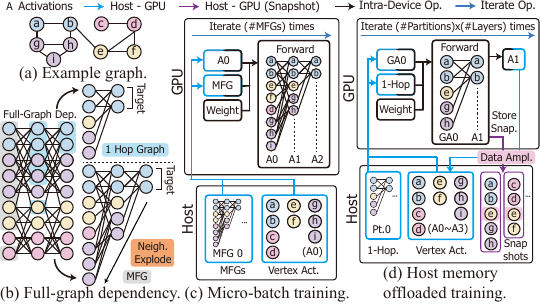}
    \vspace{-5mm}
    \caption{Full-graph training of single epoch for limited resources.}
    \label{fig:background_app}
\end{figure}

A few single-server approaches exist to address such issues.
Micro-batch~\cite{betty} and host memory offloaded~\cite{hongtu} training have tried to conduct full-graph training in GPU memory-limited environments.
\cref{fig:background_app} illustrates an example graph and discusses the drawbacks of the above methods based on the full-graph dependency.

\textbf{Micro-Batch Training}:
Betty~\cite{betty} (\cref{fig:background_app}c) accumulates gradients from message flow graphs (MFGs) with all neighbor information across all layers, followed by a single weight update.
However, even a small number of GNN layers cause MFGs to expand rapidly (\cref{fig:background_app}b), often exceeding the GPU memory.
Partitioning~\cite{metis, parmetis, mtmetis} can reduce MFG size but requires significant memory, presenting a practical bottleneck.

\textbf{Host Memory Offloaded Training}:
HongTu~\cite{hongtu} (\cref{fig:background_app}d) reduces GPU memory usage by moving activations and gradients to host memory.
A 1-hop partitioning approach extracts 1-hop graphs that fit in GPU memory.
During an epoch, activations for these graphs are transferred to the GPU, processed, and offloaded back to host memory.
While this saves GPU resources, it causes a \emph{data amplification} problem: by saving `snapshots' of 1-hop graphs for the backward pass, vertices appearing in multiple 1-hop graphs are stored repeatedly, increasing memory and I/O overhead.

\label{sec:background:partitioning}

\textbf{Memory-Hungry Partitioning}:
All aforementioned methods rely on partitioning tools like METIS~\cite{metis, parmetis, mtmetis}, which sequentially coarsen and refine graphs.
This process consumes up to 4.8$\times$ the graph’s size in memory~\cite{metis_memory}, often exceeding the capacity of a typical single server. 
Hence, existing single-server full-graph methods face out-of-memory risks or resort to an external server.

\section{Limitation of Extending Storage-Based Mini-Batch Training to Full-Graph Training with Micro-Batch Training}
\label{app:extension_limit}

While extending storage-based mini-batch training (e.g., Ginex~\cite{ginex}, MariusGNN~\cite{marius}, DiskGNN~\cite{diskgnn}, GNNDrive~\cite{gnndrive}) to full-graph training by setting the batch size to the entire node set and maximizing the neighbor size (i.e., micro-batch training from Betty~\cite{betty}) may seem to enable the large-scale full-graph training on a limited environment, it faces several limitations\footnotemark[3].

\footnotetext[3]{Please note that since these systems are not designed for full-graph training workflows, adaptations may deviate from their original performance characteristics.}

First, as it still depends on the message flow graph structure (MFG), it faces the GPU memory limit like the original micro-batch training (Betty).
Since micro-batch training needs to keep all neighbor information intact without sampling, it easily runs into out-of-memory due to neighbor explosion.
For more details, please refer to \cref{app:prev_works}.

Second, they are mainly focused on handling initial features efficiently for mini-batch training, and are inefficient in supporting full-graph training without sampling.
For instance, DiskGNN aggressively utilizes the preprocessing and pre-stores the mini-batch message flow graphs and related initial features to efficiently support large-scale mini-batch training with storage.
However, since full-graph training (with micro-batch training) needs to handle all the features without dropping, the preprocessed data size easily exceeds the SSD capacity due to the redundantly saved data.

\begin{table}[t]
\centering
\caption{\textbf{Performance of extending storage-based mini-batch training to full-graph training.} $^\dagger$: GNNDrive’s GPU caching is statically preprocessed, so the fanout is restricted to 25 and not equivalent to full-graph training. $^*$: Preprocessing failed because of excessive disk space usage.}
\label{tab:disk_baselines}
\vspace{0.1in}
\setlength{\tabcolsep}{8pt}
\resizebox{\columnwidth}{!}
{
    \begin{tabular}{lcccc}
    \toprule
    \textbf{Method} & \textbf{Products} & \textbf{IGBM} & \textbf{Papers} \\
    \midrule
    Ginex & 9.00 & OOM & 17.72 \\
    DiskGNN & 2.18 & Preproc. Fail$^*$ & Preproc. Fail$^*$ \\
    GNNDrive & 6.33$^\dagger$ & OOM & 12.06$^\dagger$ \\
    \rowcolor{gray!15}\textbf{GriNNder (Ours)} & \textbf{0.12} & \textbf{0.93} & \textbf{9.07} \\
    \bottomrule
    \end{tabular}
}
\end{table}

To show the above limitations directly, we evaluated DiskGNN~\cite{diskgnn} and GNNDrive~\cite{gnndrive}, which surpass the previous state-of-the-art storage-based mini-batch training (Ginex~\cite{ginex}, MariusGNN~\cite{marius}) in \cref{tab:disk_baselines}.
Ginex and GNNDrive encountered GPU out-of-memory on IGBM due to neighbor explosion without information dropping.
On Papers, even with fanout 25, they were significantly slower than ours.
DiskGNN uses offline preprocessing to pre-store all cacheable mini-batches with features. The preprocessing of IGBM/Papers fails by overflowing 4TB SSD, even with reduced fanout (25) from neighbor explosion.
Our method is significantly faster for runnable cases (Products/Papers).

To sum up, while mini-batch storage-based systems can emulate full-graph training by micro-batch training, results show that this becomes infeasible on a large scale.
This is due to either GPU memory exhaustion or prohibitive preprocessing disk usage.
GriNNder avoids them by not relying on message flow graphs (MFGs) or redundant preprocessing.

\section{Overall Procedure of \thiswork}
\label{app:algo}

\begin{algorithm}[t]
\caption{Overall procedure of \thiswork}
\label{alg:overall}

\begin{algorithmic}[1]
\scriptsize
 \renewcommand{\algorithmicrequire}{\textbf{Input:}}
 \REQUIRE $ $ $\{W^{i}|1\leq i \leq L\}$: initial parameters, $L$: \#layers\\
 $G$: graph,  $F$: initial features, $P$: \#partitions to meet GPU mem. req.\\
{\hspace{-6mm}\textbf{Output:}} $\{W^{i}|1\leq i \leq L\}$: updated parameters\\
{\hspace{-6mm}\textbf{Notations:}}\\
$T_{p}$: 1-hop topologies (src$\xrightarrow{}$dst)\\
$A^{l}_{p}$: destination features/activations of layer $l$, partition $T_{p}$\\
$GA^{l}_{p}$: gathered source features/activations of layer $l$, partition $T_{p}$

\STATE{\textbf{if} $~METIS_{limit} \geq Host_{limit}$ \textbf{then}}
    \STATE{$~~~T_{(\cdot)} \leftarrow \mathit{SA\_Partition (G, P)}$} {\color{blue(ncs)}{~~// \Ourpart (Sec. \ref{sec:partitioning})}}
\STATE{\textbf{else}$~T_{(\cdot)} \leftarrow \mathit{METIS (G, P)}$ \textbf{end if}}
\STATE{\color{blue(ncs)}{~~// Do until finding proper $P$ which makes all $T_{p}$s fit GPU memory limit.}}
\STATE{}

\FOR{$e=1$ ... $\#epochs$}
    \STATE{\color{blue(ncs)}{ // Forward pass}}
    \STATE{\textbf{for} $l=1$ ... $L$ \textbf{do}}
        \STATE{\;\;\;\;\;$Storage\_to\_Host(A_{(\cdot)}^{l-1})$\color{blue(ncs)}{~~// \Ourstructure (\cref{sec:caching})}}
        \STATE{\;\;\;\;\;\textbf{for} $p=0$ ... $P-1$ \textbf{do}}
            \STATE{\;\;\;\;\;\;\;\;\;\;$GA_{p}^{l-1} \leftarrow Gather(A_{(\cdot)}^{l-1})$}
            \STATE{\;\;\;\;\;\;\;\;\;\;$ Host\_to\_GPU(GA_{p}^{l-1})$}
            \STATE{\;\;\;\;\;\;\;\;\;\;$A_{p}^{l} \leftarrow FW(W^{l}, GA_{p}^{l-1}, T_{p})$ \color{blue(ncs)}{~~// w/ Regathering (redundancy elimination) (\cref{sec:ioopt})}}
            \STATE{\;\;\;\;\;\;\;\;\;\;$GPU\_to\_Host(A_{p}^{l})$}
        \STATE{\;\;\;\;\;\textbf{end for}}
    \STATE{\textbf{end for}}

    \STATE{\color{blue(ncs)}{ // Backward pass}}
    \FOR{$l=L$ ... $2$}
        \STATE{\color{blue(ncs)}{ // \Ourstructure (\cref{sec:caching})}}
        \STATE{\;\;\;\;\;$Host\_Upload\_or\_Intitialization(A_{(\cdot)}^{l-1}, \nabla A_{(\cdot)}^{l-1})$\color{blue(ncs)}{\;\;\; // Host as write-back buffer}}
        \STATE{\;\;\;\;\;\textbf{for} $p=0$ ... $P-1$ \textbf{do}}
            \STATE{\;\;\;\;\;\;\;\;\;\;$Storage\_to\_Host(A_{p}^{l}, \nabla A_{p}^{l})$}
            \STATE{\;\;\;\;\;\;\;\;\;\;$GA_{p}^{l-1} \leftarrow Gather(A_{(\cdot)}^{l-1})$}
            \STATE{\;\;\;\;\;\;\;\;\;\;$ Host\_to\_GPU(GA_{p}^{l-1})$\color{blue(ncs)}{\;\;\;\;\;\;\;// \Ourgrad (\cref{sec:ioopt})}}
            \STATE{\;\;\;\;\;\;\;\;\;\;($\nabla GA_{p}^{l-1},\nabla W^{l}$) $\xleftarrow{\cdot,+} BW(W^{l},A_{p}^{l},\nabla A_{p}^{l}, GA_{p}^{l-1})$}
            \STATE{\;\;\;\;\;\;\;\;\;\;$GPU\_to\_Host(GA_{p}^{l-1})$}
            \STATE{\;\;\;\;\;\;\;\;\;\;$\nabla A_{(\cdot)}^{l-1} \xleftarrow{+} Scatter(GA_{p}^{l-1})$}
        \ENDFOR

\ENDFOR

\end{algorithmic}
\end{algorithm}

As \thiswork is the first work on storage offloaded full-graph GNN training, we carefully designed the framework to address the three challenges outlined in \cref{sec:intro}, whose overall procedure is listed in \cref{alg:overall}.
\thiswork first partitions the graph into smaller pieces, which should be done to incur minimal data transfer (line 2). 
Our contribution is on devising a lightweight partitioning algorithm that operates with significantly lower memory requirements while preserving the partitioning quality (\cref{sec:partitioning}).
Then, for each partition (lines 10 and 21), forward and backward passes are performed on the GPUs (lines 11--14 and 22--27).
To maximize the reuse of the data, \thiswork designs an efficient policy to cache intermediate data on the host memory (\cref{sec:caching}). 
During the forward/backward passes, much of the data transfer occurs between GPU-CPU due to checkpointing (lines 12, 14, 24, 26).
This was originally designed toward reducing latency in previous work~\cite{hongtu}, but it severely increases the amount of traffic and host memory usage for storage offloading scenarios.
\thiswork redesigns the gradient engine with redundancy elimination, achieving significantly higher speedup and less memory requirement (\cref{sec:grad_engine}).

\section{Profile of Dependency among Partitions}
\label{app:partition_profile}

\begin{figure}[t]
    \centering
    \includegraphics[width=\columnwidth]{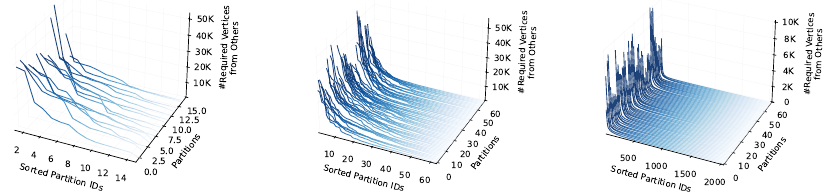}
    \caption{Partition dependency profile. (left) Products with 16 partitions, (mid) IGBM with 64 partitions, and (right) Papers with 2048 partitions. In the case of Papers, we only presented earlier 64 partitions for visibility.}
    \label{fig:part_profile}
\end{figure}

We additionally presented the profile of dependency among partitions on other datasets in \cref{fig:part_profile}.
When the size of a graph becomes larger, we need to partition the graph into a much larger number of partitions.
This makes the trend of power-law distribution clearer.
For instance, in \cref{fig:part_profile}(right), the Papers dataset with 2048 partitions shows a very vivid power-law distribution compared to the other two cases.
This further enhances the scalability of \thiswork on large-scale graphs.

\section{Vertex-Wise Cache Management vs. Partition-Wise Cache Management}
\label{app:vertex_vs_partition}

\begin{figure}[t]
    \centering
    \includegraphics[width=.9\columnwidth]{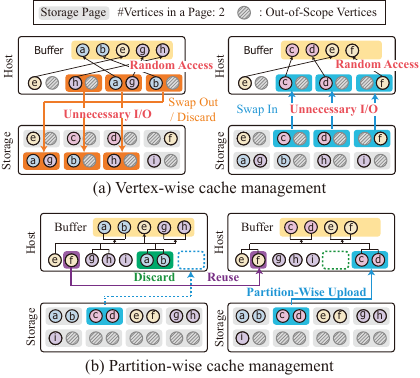}
    \caption{Advantage of partition-wise cache management compared to vertex-wise one.}
    \label{fig:vertex_vs_partition}
\end{figure}

\cref{fig:vertex_vs_partition} emphasizes the advantage of partition-wise cache management compared to the vertex-wise cache management.
Since storage devices access data at a page granularity (e.g., 16KiB), vertex-wise cache management incurs a substantial amount of \emph{unnecessary I/O}, denoted as `Out-of-Scope' vertices.
For instance, when processing the next partition, in \cref{fig:vertex_vs_partition}a, vertex-wise management needs to swap out (or discard) and swap in unnecessary data combined with the required data due to the page granularity of a storage device.
In contrast, loading and evicting at a partition granularity alleviates such overhead.

\section{I/O Optimizations of \thiswork}
\label{app:io_opt}

\subsection{Overlapping of Processing and Cache Management}

\begin{figure}[h]
    \centering
    \includegraphics[width=.8\columnwidth]{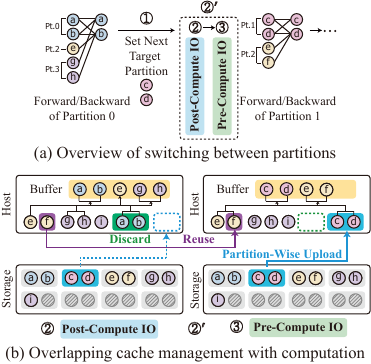}
    \caption{Overview of overlapping cache management with computation.}
    \label{fig:overlap_overview}
\end{figure}

\thiswork schedules host memory cache evictions and prefetching to overlap with GPU computations, minimizing storage I/O latency as illustrated in \cref{fig:overlap_overview}.
\circled{1} We pick the next target partition to exploit already-cached neighbors, determined statically since 1-hop graphs are fixed.
\circled{2} We discard partitions no longer needed.
\circled{3} We fetch only required partitions from storage while keeping reusable ones in the host memory.
Because we keep a small extra buffer (dotted blue), uploading the dependency for partition 1 (pre-compute) does not have to wait for partition 0 computation and the succeeding evictions (post-compute), which enables overlapping these I/O operations (\circled{2}') with ongoing computations.
Also, we overlap the GPU compute and host-GPU I/O to further reduce latency.

\begin{figure*}[t]
    \centering
    \includegraphics[width=\textwidth]{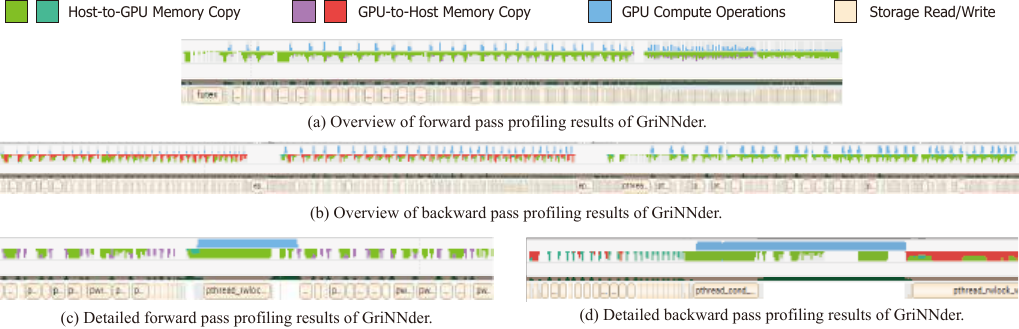}
    \caption{Profiling results of \thiswork's forward and backward pass.}
    \label{fig:io_profile}
\end{figure*}

We actually profile the training procedure of \thiswork, as illustrated in \cref{fig:io_profile}.
We profiled the 3-layer GCN on the IGBM dataset with \#partitions=32.
In both forward and backward passes, \thiswork overlaps the host memory and storage I/O with the GPU computation.
Thus, in overall training, \thiswork enables aggressive latency overlapping of I/O and computation and provides superior training throughput.

\subsection{In-Partition Vertex Ordering for Sequential Accesses}

Another source of slowdown is in the gathering, which places vertex activations to be sent ($GA$) to the GPU in a dedicated host buffer. 
This involves multiple random memory accesses, as illustrated in \cref{fig:vertex_vs_partition}a, causing slowdown.
To avoid this, after the graph is partitioned, we reorder the individual adjacency lists such that the neighbors are first sorted by their partition IDs and then by their vertex IDs. 
This replaces the random lookups with a single random lookup per partition, as in \cref{fig:vertex_vs_partition}b.

\section{In-depth I/O Volume and Memory Footprint Analyses}
\label{app:indepth_io}

\begin{table}[h]
\centering

\caption{\textbf{I/O analysis in forward pass}}
\label{tab:communication_app}
\setlength{\tabcolsep}{1pt}
\resizebox{\columnwidth}{!} {
\begin{tabular}{c|ccc}
\toprule
Methods & GPU-Host & Host-Storage & GPU-Storage \\
\midrule
\makecell{HongTu w/ OS swap memory\\(i.e., \texttt{mmap})} & $(2 \alpha + 1) D$ & $ (2 \alpha + 1) D - Mem_{Host}$ & \\
\rowcolor{gray!15}Ours & $\alpha D$ & $\alpha D - Cache Hit$ & $D$ \\

 \bottomrule
 \multicolumn{4}{r}{\small{* $|V||H|=D$. Topology data I/O is omitted for brevity.}}\\
 \multicolumn{4}{r}{\small{}}
\end{tabular}
}

\end{table}

\begin{table}[h]
\centering
\vspace{-4mm}
\caption{\textbf{Maximum memory usage analysis}}
\label{tab:memory}

\setlength{\tabcolsep}{3pt}
\resizebox{.65\columnwidth}{!} {
\begin{tabular}{c|cc}
\toprule
Methods & Host & Storage \\
\midrule
HongTu & $(\alpha+1)D|L| + 2D$ & \\
\rowcolor{gray!15}Ours & $D+D$ & $D|L|+D$\\
 \bottomrule
 \multicolumn{3}{r}{* \small{$|V||H|=D$. Considers activation and gradients.}}\\
\end{tabular}
}

\end{table}

\Ourgrad greatly reduces the I/O volume from snapshot store/load per layer and the memory footprint displayed in \cref{tab:communication_app} and \cref{tab:memory}, where $D=|V||H|$.

\textbf{I/O Volume}:
We assume host memory offloaded training (HongTu~\cite{hongtu}) to utilize OS swap memory (i.e., \texttt{mmap}), since it targets the host memory, not storage employment.
In \cref{tab:communication_app}, compared to HongTu, the input activation-related GPU-host I/O volume ($2\alpha D$) is halved ($\alpha D$) by skipping snapshots.
\thiswork incurs $\alpha D - Cache Hit$ amount of host-storage traffic for \ourpolicy, but this is significantly less than utilizing \texttt{mmap} swap memory.
Also, when host memory can handle the single-layer activations ($D$), this term becomes $D$ from a full hit.
When the host memory offloaded training faces the memory limit ($Mem_{Host}$), it needs to swap around $(2\alpha +1)D - Mem_{Host}$ data from/to storage.
Given that $\alpha$ is around 3-10, the improvement is significant.

\textbf{Memory Footprint}: In \cref{tab:memory}, we report the peak memory usage of host offloaded training~\cite{hongtu} (HongTu) and \thiswork.
For HongTu, the overhead mostly comes from storing snapshots for all layers.
These redundant snapshots consume additional $\alpha D |L|$ on top of $D|L|$ activations.
It needs to save $2D$ of gradients in backward pass to handle input and output gradients.
In contrast, with \ourgrad and \ourstructure, \thiswork consumes up to $D + D$ host memory for saving layer-wise activations and gradients.
Regarding storage usage, \thiswork consumes $D|L|$ for saving activations and $D$ for single-layer gradients.

\section{Insights and Details of \OurPart}
\label{app:part}

We draw inspiration from streaming partitioning (Spinner~\cite{spinner}), which applied traditional label propagation~\cite{lpa} to partitioning in distributed cloud graph systems (e.g., Pregel~\cite{pregel}).
While lightweight label propagation suits our host memory constraints, Spinner's message-passing-based design is unsuitable for such limited environments.

Hence, we propose \ourpart, which adapts label propagation for limited resources with memory usage similar to CSR.
We also introduce a group-wise propagation strategy suited for storage-offloaded full-graph training. 

\Ourpart aims to find vertices with similar properties in different partitions and relocate them to the same partition. 
Additionally, we need to balance the size of each partition to reduce the workload imbalance between partitions.
To do so, we iteratively refine the partitions by selectively relocating vertices within a certain limit.

\begin{figure}[t]
    \centering
    \includegraphics[width=\columnwidth]{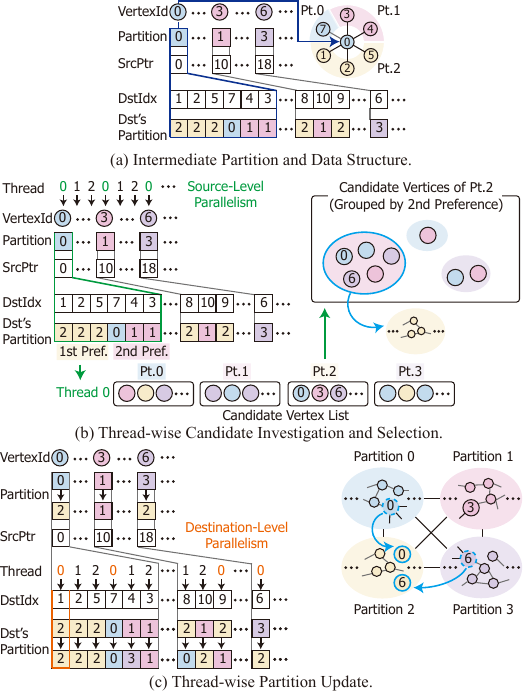}
    \vspace{-7mm}
    \caption{\Ourpart.
    }
    \label{fig:partition_app}
\end{figure}

\cref{fig:partition_app} shows the detailed procedure of the proposed \ourpart.
At first, we set the initial partitioning state ($S_{0} = {P_{0}, ..., P_{p-1}}$) by randomly assigning each vertex to different partitions. 
We want to achieve high-quality partitioning while maintaining the number of vertices of all partitions close to $|V|/p$. 
$|\cdot|$ means the \#vertices in a partition (or a graph), and $p$ is the \#partitions.
We additionally define the maximum capacity term as $\beta$ and set the maximum capacity limit of a single partition as $\beta \times |V|/p$.
Here the capacity of a partition refers to the number of vertices allocated to said partition.
In a state $S_{i}$, each partition $j$ has the available relocation capacity ($RC_{(i,j)}$) as follows:
\begin{equation}
\begin{aligned}
RC_{(i,j)} = \beta \times |V|/p - |P_{j}|, (0 \leq j < p)
\label{eq:background}
\end{aligned}
\end{equation}
This is used to limit the number of vertices moved to the current partition.  
\cref{fig:partition_app}a illustrates the intermediate state ($S_{i}$) where each partition has the available relocation capacity of six ($RC_{(i,j)}=6$).
Following the CSR format, our data structure comprises source pointers (\texttt{SrcPtr}) and destination indices (\texttt{DstIdx}).
We manage another array (\texttt{Dst's Partition}) and fill this array with the partition of each destination index in \texttt{DstIdx}.
For example in \cref{fig:partition_app}b, the vertex $0$ has neighbors of vertex $\{1,2,5,7,4,3\}$.
For each neighbor, we fill the \texttt{Dst's Partition} with its partition $\{2,2,2,0,1,1\}$.

\begin{sloppypar}
From a state ($S_{i}$) (\cref{fig:partition_app}a),
we calculate \textit{$k$th preference} of a vertex: among the neighbors of the vertex, the partition ID of the $k$th largest frequency is the $k$th preference of the vertex.
Then, using each vertex's first preference, each partition manages its own relocation candidate vertices from other partitions. 
In \cref{fig:partition_app}b, vertex $0$'s neighboring vertices' partitions are $\{2,2,2,0,1,1\}$.
Among them, the partition that occurs most frequently is $2$.
Therefore, we put vertex $0$ to the partition $2$'s relocation candidate ($0$ is now included in Pt.2 List in \cref{fig:partition_app}b).

When selecting the final vertices to be relocated among candidates, we select them in a group-wise manner.
In \cref{fig:partition_app}b, we first use the 2nd preference partition of each vertex as a feature to help cluster vertices into different groups, unlike the baseline streaming partitioning algorithm.
We then choose the largest group with the same 2nd preference to avoid vertices belonging to small, disparate clusters being relocated.
In this example, vertex $0$'s 2nd preference partition is partition $1$, and vertex $6$'s 2nd preference partition is also partition $1$.
Therefore, we put those two vertices into the same group.
We choose to relocate the group including $\{0,6\}$ because it is the largest group among the candidates.
This provides a clustering effect and helps the convergence speed of partitioning.
This can be generalized into comparing until $k$th preference, but we use $k=2$ as default because it already empirically provides good performance.

To parallelize the procedure, we apply source-level parallelism, which distributes the source vertices to each thread.
Each thread manages its own candidate lists for partitions as depicted in \cref{fig:partition_app}b with the example of thread $0$.
We dedicate each thread to the equal available relocation capacity ($RC_{(i,j)}/\#threads$) to run threads in a fully parallel manner.
\end{sloppypar}

After selection, we update the relocation result to the \texttt{Dst's Partition} array.
In \cref{fig:partition_app}c, vertex $0$ and $6$ are selected to be relocated to partition $2$.
Therefore, we update the values of vertex $0$ and $6$ in \texttt{Dst's Partition} array to $2$ (meaning partition $2$).
This procedure is conducted with destination-level parallelism, as illustrated in \cref{fig:partition_app}c.
After the update, using the updated data structure, iteration $i+1$ proceeds.
For each iteration $i$, we conduct the following procedure until reaching the termination condition, which will be discussed in the next subsection.

We discussed \ourpart as a procedural view.
In the detailed algorithm, we need a penalty term for suppressing the propagation to reduce the imbalance among the number of vertices in partitions.
Therefore, in a state $S_{i}$, for a vertex $v$, the scoring term ($Score_{(v,I,j)}$) for each partition $j$ and the final objective are as follows:
\begin{equation}
\begin{aligned}
Penalty_{(i,j)} &= |P_{j}|/(\alpha_{balance} \times |V|/p), (0 \leq j < p)\\
Score_{(v, i, j)} &= 1 + \#N(v,j)/\#N(v,\cdot) - Penalty_{(i,j)}\\
max&imize \textstyle \sum_{v \in G} {Score_{(v,i,j=partition_{v})}}\\
\end{aligned}
\end{equation}
where $\#N(v,j)$ denotes the frequency of partition $j$ among the neighbors of the vertex $v$ and $Penalty_{(i,j)}$ denotes the penalty term of state $S_i$ of partition $j$.
The penalty term reduces the preference when a partition already reaches the additional capacity $\alpha_{balance}$.
The objective function calculates the total sum of the internal preferential scores of partitions.
The partitioning halts when the objective does not improve over $\epsilon=0.001$ for $w=5$ times.

\section{Effect of Partitioning on Host$\leftrightarrow$GPU Traffic}
\label{app:effect_partitioning}

\begin{table}[h]
\centering
\caption{\textbf{Training time breakdown of storage-offloaded training.} Storage and Host$\leftrightarrow$GPU denote the storage I/O and host$\leftrightarrow$GPU I/O.}
\label{tab:breakdown}
    \vspace{0.1in}
    \setlength{\tabcolsep}{2.5pt}
    \resizebox{\columnwidth}{!}
    {
\begin{tabular}{lccccc}
\toprule
 & \textbf{Storage} & \textbf{Host$\leftrightarrow$GPU} & \textbf{Compute} & \textbf{Sync.} & \textbf{Etc.} \\
\midrule
\textit{w \thiswork} & 19.2\% & 48.4\% & 8.6\% & 12.0\% & 11.8\% \\
\textit{w/o \thiswork} & 85.4\% & --- & 2.1\% & \multicolumn{2}{c}{12.5\%} \\
\bottomrule
\end{tabular}
}
\end{table}

Partitioning is significantly helpful in reducing host $\leftrightarrow$ GPU traffic.
Once partition-wise scheduling reduces random storage I/O and host-memory caching minimizes storage traffic, host $\leftrightarrow$ GPU traffic becomes the main bottleneck as illustrated in \cref{tab:breakdown}.
We profiled a single epoch of storage-offloaded training of the IGBM dataset without/with GriNNder's optimizations.
Since GriNNder is highly optimized to overlap compute and I/O, we employed a CPI-stacking-like method~\cite{cpi_stack} to roughly break down execution time. 
When applying GriNNder optimizations, reduced random storage accesses from partition-wise I/O and minimized storage I/O from host memory caching make the host-to-GPU traffic the main bottleneck.

Without effective partitioning, the input activations required per partition---scaled by the expansion ratio ($\alpha$)---would dramatically increase host $\leftrightarrow$ GPU traffic, severely harming performance.
In detail, for every vertex that belongs to a partition, we need to fetch all dependent extra vertices through the host $\leftrightarrow$ GPU path redundantly.
The factor that affects the \textit{extra fetching} is our expansion ratio ($\alpha$),
\begin{equation}
\alpha = \frac{\text{\# required vertices}}{\text{\# target vertices}}.
\end{equation}
Put differently, $\alpha$ tells us how much larger the input tensor (to be sent between host $\leftrightarrow$ GPU) becomes after we consider dependencies.
The total input-activation traffic for one partition is then
\begin{equation}
\text{Traffic} = \alpha \times |H| \times |N_{target}|,
\end{equation}
where $|H|$ is the hidden dimension and $|N_{target}|$ is the number of vertices the partition owns.
Partitioning is designed to drive down $\alpha$ by co-locating vertices that are frequently accessed together as much as possible.

In practice, this helps \thiswork to provide stable training speed, even when long-tail effects may cause some partition dependencies to span multiple partitions.
For instance, on the Papers dataset (2048 partitions), a target partition depends on 1602.07± 234.11 (Std.) other partitions.
With the effect above, \thiswork provides significant speedup over other baselines by reducing storage I/O and the traffic between host and GPU.

\section{API Interface Details}
\label{app:api}

\begin{figure}[h]
    \centering
    \includegraphics[width=\columnwidth]{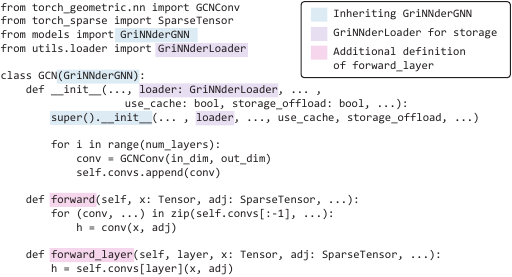}
    \vspace{-3mm}
    \caption{User interface of \thiswork. 
    Users inherit \texttt{GriNNderGNN} and implement \texttt{layer\_forward} 
    to enable layer-wise execution for partition-based full-graph training.}
    \label{fig:api}
\end{figure}

\cref{fig:api} illustrates the user-facing API design.
Existing PyG models require minimal changes: users inherit \texttt{GriNNderGNN} 
and implement the \texttt{layer\_forward} method, which receives a partition ID 
and returns the layer's output for that partition.
This enables the framework to orchestrate partition-wise execution transparently 
while maintaining compatibility with standard PyTorch training loops.

\section{Detailed Experimental Settings and Baselines}
\label{app:setup}

\begin{table}[h]
    \centering
    \caption{
    \textbf{Real-world graph datasets and hyper-parameters}
    }
    \label{tab:datasets}
    \vspace{0.1in}
    \setlength{\tabcolsep}{2pt}
    \resizebox{\columnwidth}{!}
    {

    \begin{tabular}{ccccccc}
    \toprule
         \multirowcell{2}{\makecell{Name}}
         &\multicolumn{3}{c}{Dataset Info.}
         & \multicolumn{3}{c}{Hyper-parameter} \\
         \cmidrule(lr){2-4}
         \cmidrule(lr){5-7}
         & \makecell{\#Nodes}
         & \makecell{\#Edges}
         & \makecell{Feat. size}
         &\makecell{lr}
         & \makecell{Dropout}
         & \makecell{\#Epochs} \\
          \midrule
          Products~\cite{ogb}
          & 2,449K
          & 61.9M
          & 100 & 0.003 & 0.3 & 500 \\

          IGBM~\cite{igb}
          & 10,000K
          & 120.1M
          & 1024 & 0.01 & 0.5 & 500 \\

          Papers~\cite{ogb}
          & 111,000K
          & 1,600M
          & 128 & 0.01 & 0.5 & 500 \\

        \bottomrule
    \end{tabular}
    }
\end{table}

\noindent\textbf{Models and datasets.} We tested graph convolutional network (GCN) as the baseline GNN architecture and also used GAT~\cite{gat} and GraphSAGE~\cite{graphsage}.
We set the hidden size as the widely-used 256, if not stated otherwise.
We used three medium- (Products~\cite{ogb}) to large-scale (IGBM~\cite{igb} and Papers~\cite{ogb}) datasets (details in \cref{tab:datasets}).
Products is a co-purchasing network where vertices represent Amazon products and edges indicate products purchased together.
IGBM and Papers are citation networks with vertices and edges representing research papers and citations, respectively.
We also utilized Kronecker random graphs~\cite{kronecker} (average degree=10) with random initial features of dimension 128 and \#classes of 10 for scalability and versatility tests with ablation.

\textbf{Hardware.} 
We run main experiments on a single GPU workstation with an AMD Ryzen9 7950X 3D CPU (16C 32T), 128GB DDR5-5600 RAM, one RTX A5000 (24GB) GPU, a PCIe 5.0 NVMe SSD (4TB), and a total 4TB swap space for swap-based experiments.
We utilized the NVMe SSD for the swap memory and GPUDirect Storage (GDS)~\cite{gds} and AIO~\cite{aio}.
We chose a single GPU setup to demonstrate how \thiswork breaks through the host/GPU memory limitations.
For the multi-GPU extension, we utilized a multi-GPU server with four RTX4090 GPUs, 2$\times$Intel Xeon Gold 6442Y, 512GB DDR5 DRAM, and 2TB PCIe5.0 NVMe SSD.
We used a four-server cluster to test distributed full-graph training baselines, each server having four RTX A6000 GPUs, which aggregates to 16 GPUs.
Intra-server GPUs are connected via NVLink Bridge~\cite{nvbridge}, and servers are connected via Infiniband SDR~\cite{infiniband}.
Each server has 512GB DDR4 RAM and an EPYC 7302 (16C 32T).
For IGBM/Papers, we needed all 16 GPUs to fit the data in the GPU memory.
For Products, using fewer GPUs could yield better performance, but we used all GPUs to maintain consistency among datasets.

\begin{sloppypar}
\noindent\textbf{Baselines.}
We compared four single-server baselines with \thiswork (denoted as `GRD').
For \micro full-graph training, we used Betty~\cite{betty} (called micro-batch training), the state-of-the-art full-graph training in limited environments, as our baseline.
As Betty sometimes shows significant slowdowns due to slow MFG generation, we excluded the MFG generation time for comparison.
To test extension of storage-based mini-batch training to full-graph training while utilizing SSD, we extend Ginex~\cite{ginex} to micro-batch training~\cite{betty}.
For host offloaded full-graph training, we faithfully implemented HongTu~\cite{hongtu} and used it as a baseline.
When the training data overflows the host memory, we use storage swap memory to compare it with \thiswork regarding storage usage.
We also tested the \naive extension of ROC~\cite{roc} to \naively just use storage for offloading, but reported the results of it only in \cref{app:naive_storage} because this extension was much slower than the others.
In the appendix, we additionally tested two storage-based mini-batch training (DiskGNN~\cite{diskgnn} and GNNDrive~\cite{gnndrive}) with micro-batch extension (\cref{app:extension_limit}).
\end{sloppypar}

We also compared \thiswork with two distributed full-graph training baselines, CAGNET~\cite{cagnet} and Sancus~\cite{sancus}.
CAGNET is one of the famous distributed full-graph training methods, and Sancus accelerated it by storing stale activations and gradients to reduce the communication bottleneck.
Note that while Sancus is not the exact full-graph training from using stale activations and gradients, we still included it as it is one of the state-of-the-art distributed full-graph training frameworks.
These two baselines ran on the cluster mentioned above.
When GPU out-of-memory issues arise in distributed training baselines, we implement host memory activation checkpointing (indicated by `*') to attempt to make them executable.

For partitioning, we utilized the multi-threaded METIS (MT-METIS)~\cite{mtmetis} as the baseline, which is one of the state-of-the-art METIS parallelizations (denoted as `METIS'). 
Even when it does not run on the testbed due to insufficient memory, we assume it was preprocessed in another environment since all baseline methods rely on METIS.
For comparisons with lightweight partitioners, we benchmarked Spinner~\cite{spinner} and an out-of-core partitioner (2PS-L~\cite{2ps_l}).

\section{Comprehensive Analysis with Synthetic Graph on Scalability, Ablation, and Configuration}
\label{app:synth}

\begin{table}[h]
    \centering
    \caption{
    \textbf{Training time/epoch (min) for various-sized Kronecker synthetic graphs.}
    `-' denotes when the number of partitions is not enough for running.
    \textbf{Bold} is the fastest training time in each (\#layers, dataset) pair.
    }
    \vspace{0.1in}
    \setlength{\tabcolsep}{7pt}
    \resizebox{\columnwidth}{!}{
        \begin{tabular}{lllcccc}
            \toprule
            \#Layers & \#Partitions & Method & 4.2M & 8.4M & 16.8M & 33.6M \\
            \midrule
            \multirow{12}{*}{3} 
              & \multirow{3}{*}{16}  & HongTu & 0.43 & 0.83 & - & - \\
              &                       & GRD-G  & \cellcolor{gray!15}\textbf{0.29} & \cellcolor{gray!15}\textbf{0.59} & - & - \\
              &                       & GRD-GC & 0.31 & 0.63 & - & - \\
              \cmidrule(lr){2-7}
              & \multirow{3}{*}{32}  & HongTu & 0.57 & 1.11 & 7.25 & - \\
              &                       & GRD-G  & 0.31 & 0.66 & - & - \\
              &                       & GRD-GC & 0.33 & 0.71 & - & - \\
              \cmidrule(lr){2-7}
              & \multirow{3}{*}{64}  & HongTu & 0.76 & 1.76 & 10.70 & - \\
              &                       & GRD-G  & 0.41 & 0.77 & - & - \\
              &                       & GRD-GC & 0.43 & 0.81 & - & - \\
              \cmidrule(lr){2-7}
              & \multirow{3}{*}{128} & HongTu & 1.05 & 5.32 & 18.96 & 36.31 \\
              &                       & GRD-G  & 0.55 & 1.02 & \cellcolor{gray!15}\textbf{1.93} & \cellcolor{gray!15}\textbf{3.73} \\
              &                       & GRD-GC & 0.58 & 1.05 & 1.99 & 3.86 \\

            \cmidrule(lr){1-7}
            \multirow{12}{*}{5} 
              & \multirow{3}{*}{16}  & HongTu & 0.83 & 1.99 & - & - \\
              &                       & GRD-G  & \cellcolor{gray!15}\textbf{0.57} & \cellcolor{gray!15}\textbf{1.14} & - & - \\
              &                       & GRD-GC & 0.60 & 1.20 & - & - \\
              \cmidrule(lr){2-7}
              & \multirow{3}{*}{32}  & HongTu & 1.07 & 8.04 & 19.15 & - \\
              &                       & GRD-G  & 0.60 & 1.30 & - & - \\
              &                       & GRD-GC & 0.63 & 1.37 & - & - \\
              \cmidrule(lr){2-7}
              & \multirow{3}{*}{64}  & HongTu & 1.48 & 11.43 & 24.08 & - \\
              &                       & GRD-G  & 0.79 & 1.49 & - & - \\
              &                       & GRD-GC & 0.84 & 1.55 & - & - \\
              \cmidrule(lr){2-7}
              & \multirow{3}{*}{128} & HongTu & 4.61 & 17.08 & 37.09 & 96.99 \\
              &                       & GRD-G  & 1.08 & 1.96 & \cellcolor{gray!15}\textbf{3.71} & 10.87 \\
              &                       & GRD-GC & 1.13 & 2.02 & 3.82 & \cellcolor{gray!15}\textbf{7.76} \\
            \bottomrule
        \end{tabular}
    }
    \label{tab:synth_full_table}
\end{table}

We conducted a comprehensive analysis using synthetic graphs, as summarized in \cref{tab:synth_full_table}.
The tests utilized Kronecker synthetic graphs~\cite{kronecker} with sizes ranging from $2^{22}$ to $2^{25}$ nodes (4.2--33.6M) and an average degree of 10.

Across all combinations of layers and datasets, all ablations of \thiswork consistently achieved significant speedups over HongTu.
For smaller datasets, where host memory can store all intermediate activations and gradients, the configuration using only \ourgrad (`GRD-G') generally outperforms the storage-enabled version (`GRD-GC'), primarily due to cache management overhead.
However, for larger datasets, employing storage alleviates host memory cache pressure, allowing the storage-based configuration (`GRD-GC') to deliver substantial speedups over both HongTu and GRD-G.

These results demonstrate that \thiswork is highly scalable for large datasets, with storage utilization being an effective strategy for handling large graphs on a single GPU. We also observed that \thiswork occasionally requires a larger number of partitions (i.e., different configurations) than HongTu.
This is due to the GPU memory overhead introduced by overlapping GDS operations and computation.
Despite this, \thiswork continues to deliver significant performance improvements over HongTu. 
It is also important to note that the number of partitions is merely a configuration hyperparameter, and users are not burdened by the need to manually handle this difference.

\section{Cache Hit Rates}
\label{app:cache_hit}

\begin{table}[h]
    \centering
    \caption{\textbf{Cache hit rate.}}
    \label{tab:cache_hit_rate}
    \vspace{0.1in}
    \setlength{\tabcolsep}{1pt}
    \resizebox{\columnwidth}{!}{%
    \begin{tabular}{lccccccc}
        \toprule
         & \textbf{Products} & \textbf{IGBM} & \textbf{Papers} & \textbf{kron-4.2M} & \textbf{kron-8.4M} & \textbf{kron-16.8M} & \textbf{kron-33.6M} \\
        \midrule
        Hit (\%) & 28.57 & 53.70 & 83.63 & 80.81 & 80.47 & 92.77 & 92.70 \\
        \bottomrule
    \end{tabular}%
    }
\end{table}

We report cache hit rates in \cref{tab:cache_hit_rate}. As larger datasets ($>$ IGBM, 10M) incur more reuse from the higher number of partitions, the hit rate is more significant in them. A low hit rate is natural in small datasets (e.g., Products) because we employ only a few partitions, and most data are not reused. Thus, GriNNder’s caching is promising in large-graph training.

\section{Convergence Trend and Practical Overhead of \OurPart}
\label{app:part_converge}

\begin{table}[h]
    \centering
    \caption{\textbf{Partitioning convergence trend.}}
    \label{tab:partitioning_convergence}
    \vspace{0.1in}
    \setlength{\tabcolsep}{.5pt}
    \resizebox{\columnwidth}{!}{%
    \begin{tabular}{llcccccccccccc}
        \toprule
        \textbf{Dataset} & & \multicolumn{11}{c}{\textbf{Improvement (\%) for Iterations}} \\
        \midrule
        \multirow{2}{*}{\makecell{Products\\(4 parts)}} & Iter. & 1 & 5 & 10 & 15 & 20 & 25 & 28 (last) \\
        & Improve (\%) & 6.81 & 9.75 & 3.79 & 0.36 & 0.12 & 0.08 & 0.05 \\
        \midrule
        \multirow{2}{*}{\makecell{IGBM\\(32 parts)}} & Iter. & 1 & 5 & 10 & 15 & 20 & 25 & 30 & 35 & 40 & 45 & 50 (last) \\
        & Improve (\%) & 11.13 & 7.78 & 3.66 & 1.96 & 0.66 & 0.77 & 0.39 & 0.21 & 0.16 & 0.10 & 0.08 \\
        \midrule
        \multirow{2}{*}{\makecell{Papers\\(2K parts)}} & Iter. & 1 & 5 & 10 & 15 & 20 & 25 & 30 & 35 & 40 & 45 & 50 (last) \\
        & Improve (\%) & 18.04 & 2.86 & 3.96 & 1.61 & 1.78 & 0.89 & 0.46 & 0.72 & 0.43 & 0.22 & 0.14 \\
        \bottomrule
    \end{tabular}%
    }
\end{table}

\Ourpart converges fast with low practical overhead.
In \cref{tab:partitioning_convergence}, we report the trend of the partitioning quality (score of the objective function) improvement (convergence) from the adjacent previous iteration (e.g., iter 4 $\xrightarrow[]{}$ 5). We observe that at most 50 iterations are enough for convergence, thus limiting partitioning to 50 iterations in our experiments.

Given that a single iteration takes 0.08sec/0.14sec/21.12sec on average and our lightweight partitioning only requires 2.49sec/6.96sec/17.60min, partitioning consumes 0.07/0.02/0.39\% of the total training time (500 epochs) on Products/IGBM/Papers, respectively.

\section{Multi-GPU Extension}
\label{app:multi_gpu}

 Our multi-GPU implementation employs two lightweight mechanisms:

(i) Partition parallelism: We divide the partitions into disjoint \#(GPU) sets; each GPU performs forward/backward on its set independently.
(ii) Weight/Gradient synchronization: During the backward pass, partial gradients of dependent vertices from different GPUs are atomically accumulated on the host, which accumulates the gradients of the vertices. Before the weight update, a weight gradient all-reduce operation is conducted between GPUs to synchronize the weights among them.

These straightforward enhancements enable effective multi-GPU execution with minimal overhead. 

\section{Configuration Sensitivity Results}
\label{app:config_sensi}

\begin{table}[!h]
  \centering
  \caption{\textbf{Configuration sensitivity on training time (sec).} The default number of partitions for \textsc{Products} and \textsc{IGBM} are 2 and 32, respectively.}
  \label{tab:config_sensi_app}
  \vspace{0.1in}
  \setlength{\tabcolsep}{1pt}
  \resizebox{\columnwidth}{!}{%
    \begin{tabular}{lllcccc}
      \toprule
      &  & ~~Method & ~~~~~~~~\footnotesize{$\times 1$}~~~~~~~~ & ~~~~~~~~\footnotesize{$\times 2$}~~~~~~~~ & ~~~~~~~~\footnotesize{$\times 4$}~~~~~~~~ & ~~~~~~~~\footnotesize{$\times 8$}~~~~~~~~ \\
      \midrule
      \mr{4}{\rotatebox{90}{\small{3-layer}}}
       & ~\multirow{2}{*}{\textsc{Products}} & ~~HongTu & 9.98 & 11.11 & 12.22 & 13.65 \\
       &                                      & \cellcolor{gray!15}~~\textbf{GRD}    & \cellcolor{gray!15}\textbf{6.93} & \cellcolor{gray!15}\textbf{7.72}  & \cellcolor{gray!15}\textbf{8.55}  & \cellcolor{gray!15}\textbf{8.99}  \\
      \cmidrule(lr){2-7}
       & ~\multirow{2}{*}{\textsc{IGBM}}      & ~~HongTu & 387.68 & 694.02 & 675.98 & 876.60 \\
       &                                      & \cellcolor{gray!15}~~\textbf{GRD}    & \cellcolor{gray!15}\textbf{55.62}  & \cellcolor{gray!15}\textbf{59.41}  & \cellcolor{gray!15}\textbf{61.06}  & \cellcolor{gray!15}\textbf{66.39}  \\
       \midrule
        \mr{4}{\rotatebox{90}{\small{5-layer}}}
       & ~\multirow{2}{*}{\textsc{Products}} & ~~HongTu & 19.14 & 21.46 & 23.42 & 26.22 \\
       &                                      & \cellcolor{gray!15}~~\textbf{GRD}    & \cellcolor{gray!15}\textbf{13.65} & \cellcolor{gray!15}\textbf{15.22} & \cellcolor{gray!15}\textbf{16.38} & \cellcolor{gray!15}\textbf{17.60} \\
      \cmidrule(lr){2-7}
       & ~\multirow{2}{*}{\textsc{IGBM}}      & ~~HongTu & 894.09 & 958.20 & 1183.88 & 1425.36 \\
       &                                      & \cellcolor{gray!15}~~\textbf{GRD}    & \cellcolor{gray!15}\textbf{91.46}  & \cellcolor{gray!15}\textbf{92.60}  & \cellcolor{gray!15}\textbf{98.99}   & \cellcolor{gray!15}\textbf{114.76}  \\
      \bottomrule
    \end{tabular}
  }
\end{table}

We additionally conducted configuration sensitivity experiments in \cref{tab:config_sensi_app}.
From the efficient caching management and elimination of redundancy, \thiswork is much less sensitive to the number of partitions (configurations).
This enhances the practicality of \thiswork for end-users as they are not required to carefully configure the number of partitions.

\section{Heterogeneous GNN Extension}
\label{app:hetero}

We extended GriNNder to support heterogeneous graph training.
Our implementation involved creating a heterogeneous graph dataloader (HeteroDataloader).
We validated this extension using the IGBM-hetero dataset with a two-layer heterogeneous convolutional model, consisting of a GCN layer (paper-cite-paper relation) and GraphSAGE layers (other relations).
With hidden dimensions set to 128 (except for the output layer with 19 classes), GriNNder achieved 71.58\% accuracy after 100 epochs, while significantly reducing runtime compared to HongTu (26.92 sec/epoch vs. 55.41 sec/epoch).

\section{Benchmarking w/o GDS}
\label{app:gds}

GriNNder can be generally used when GDS is unavailable. In this case, Kvikio (used in GriNNder) automatically switches to POSIX. Thus, users can still utilize GriNNder without any modification. Also, please note that GDS is supported on GPUs with NVIDIA compute capability  >6.\texttt{x} (e.g., V100 and after).

\begin{table}[h]
\centering
\caption{\textbf{Sensitivity to GDS.}}
\label{tab:gds}
\vspace{0.1in}
    \setlength{\tabcolsep}{15pt}
    \resizebox{\columnwidth}{!}{%
\begin{tabular}{llccc}
\toprule
\textbf{Layers} & \textbf{GDS} & \textbf{Products} & \textbf{IGBM} & \textbf{Papers} \\ 
\midrule
\multirow{2}{*}{3 layer} & GDS      & 0.12 & 0.93 & 9.07  \\ 
                         & w/o GDS  & 0.12 & 0.93 & 10.25 \\ 
\midrule
\multirow{2}{*}{5 layer} & GDS      & 0.23 & 1.52 & 12.03 \\ 
                         & w/o GDS  & 0.23 & 1.52 & 13.73 \\ 
\bottomrule
\end{tabular}
}
\end{table}

We also benchmarked the performance (min) of GriNNder without GDS support in \cref{tab:gds} as `w/o GDS'. As Products and IGBM can be handled with host memory, the `w/o GDS' performs similarly to the GDS cases. Even with Papers, where storage is highly utilized, there is only a 13-14\% slowdown, demonstrating GriNNder’s versatility.

\section{Effect of Graph Distributions}
While power-law degree distributions are widely observed in real-world large-scale graphs~\cite{leskovec}, we also evaluate the robustness of \thiswork on non-power-law graphs.
\thiswork's partitioning and caching strategies operate independently of specific graph structural properties, enabling performance gains across diverse graph patterns.
To evaluate behavior on non-power-law graphs, we benchmark HongTu and \thiswork on a 10M Watts-Strogatz graph~\cite{watts_strogatz} (average degree=16) with randomized features using GCN in \cref{tab:non-power-law}.
\thiswork achieves 3.21$\times$/7.01$\times$ speedups for 3-/5-layer models, demonstrating robustness beyond power-law graphs.

\begin{table}[h]
\centering
\caption{\textbf{Performance on non-power-law graph.}}
\label{tab:non-power-law}
\vspace{0.1in}
\setlength{\tabcolsep}{15pt}
\resizebox{\columnwidth}{!}{%
\begin{tabular}{lccc}
\toprule
Time/Epoch (min) & HongTu & \textbf{GriNNder} & Speedup \\
\midrule
3-Layer & 3.50 & \cellcolor{gray!15}\textbf{1.09} & 3.21$\times$ \\
5-Layer & 14.72 & \cellcolor{gray!15}\textbf{2.10} & 7.01$\times$ \\
\bottomrule
\end{tabular}%
}
\end{table}

\section{Application to Dynamic Graphs}
\thiswork can be extended to dynamic graphs by modifying the data-loading and partitioning components for graph updates.
\Ourpart is especially well-suited for this: its streaming nature enables effective adaptation to incremental changes by starting from existing partitions and converging to high-quality solutions in a few lightweight iterations.
Thus, we expect \thiswork to handle dynamic graphs without significant overhead.

\section{Approaches to Resemble Full-graph Training with Algorithm Change}

Many works have been proposed to resemble the accuracy (effect) of full-graph training by addressing the information loss of mini-batch training.
GNNAutoScale~\cite{gnnautoscale} utilizes stale activation to compensate for the information loss of mini-batch training.
LMC~\cite{lmc} further addresses the information loss by compensating for it with gradients.
In distributed full-graph training, many researchers have tried to address the communication bottleneck while resembling the full-graph training accuracy with staleness~\cite{pipegcn, sancus} and error compensation~\cite{bns_gcn} through proportional dropping of communication.
While the above compensation methods could be orthogonally applied to further enhance the performance of \thiswork, we did not apply them to implement the exact full-graph training without algorithm change.

\section{Functionality (Accuracy) Check of \thiswork}
\label{sec:accuracy}

\begin{figure}[h]
    \centering
    \includegraphics[width=\columnwidth]{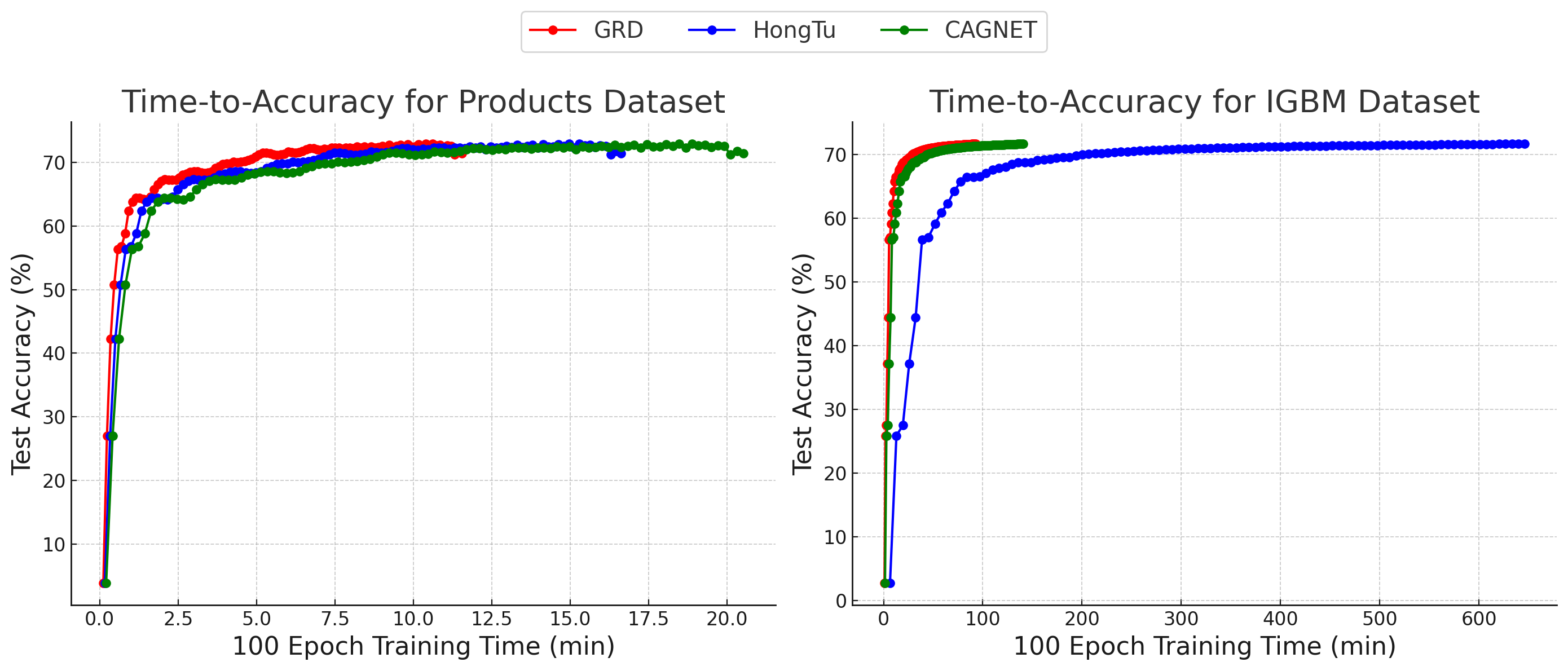}
    \caption{Functionality check of \thiswork.}
    \label{fig:acc_check}
\end{figure}

While \thiswork does not change the algorithm of full-graph training, we tested the accuracy of \thiswork compared to full-graph training and HongTu for the functionality check, as illustrated in \cref{fig:acc_check}.
Full-graph training was conducted with a CAGNET distributed baseline because Sancus is not exact full-graph training.
As depicted in \cref{fig:acc_check}, while HongTu is much slower than the distributed setup, \thiswork provides significant speedup over the distributed CAGNET.
Both baselines and \thiswork show the same accuracy, which demonstrates the correct functionality of \thiswork.
We also additionally checked Sancus's result, which is not exact full-graph training as it utilizes stale activations and gradients.
It shows similar accuracy to others but not exactly the same as them.
On the Papers dataset, GriNNder achieves identical accuracy (63.04\%) as CAGNET. 
We excluded the Papers in the figure because HongTu consistently encountered out-of-memory issues.

\section{Comparison with \Naive Baseline (\Naive Storage Extension of ROC~\cite{roc})}
\label{app:naive_storage}
We also tested the \naive storage extension of ROC~\cite{roc} instead of HongTu, which is the state-of-the-art framework.
While we tested with HongTu with OS-based swap (i.e., \texttt{mmap}), we made it directly utilize storage instead of OS-based management for the ROC extension.
On this \naive extension, \thiswork provides 1.28/29.00$\times$ speedup on 3-layer GCN on Products and IGBM, respectively.
The speedup is significant on IGBM because Products only use \#partitions=2 while IGBM uses \#partitions=32.
Thus, \thiswork provides further speedup on IGBM, which has much redundancy issue with ROC.

\section{Limitation}

\label{app:limit}

We evaluate an extensive set of datasets and demonstrate the effectiveness of our partition-wise cache management.
We also found that partition-wise cache management remains effective even when long-tail effects may cause some partition dependencies to span multiple partitions (\cref{app:effect_partitioning}).
However, there can be a worst-case scenario: when dependencies are uniformly distributed across many partitions.
In this case, partition-wise management may lead to overhead rather than performance improvement.
We leave the handling of such a case to future work.

\section{Additional Related Work}
\label{app:other_related_work}

This section provides comprehensive discussion of related work beyond the core contributions in \cref{sec:related_work}.

\noindent\textbf{Full-Graph GNN Training Systems.}
Numerous methods have been proposed for learning representations from graphs~\cite{ogb, web-scale-graph-recommendation, friend_scale, sign}.
Full-graph training preserves complete input information, preferred for algorithmic validation~\cite{pipegcn, roc, bns_gcn, neugraph}.
Distributed approaches~\cite{sancus, cagnet, distmemoryfg, g3, bgl, neutronstar} partition graphs across machines but incur substantial inter-device communication overhead.
Single-server methods enable full-graph training without distribution costs.
Betty~\cite{betty} employs batch-level graph partitioning but suffers from neighbor explosion across layers.
HongTu~\cite{hongtu} stores activations/gradients to host memory but remains limited by host capacity.
Hardware acceleration through near-memory processing~\cite{gnnear}, computational storage~\cite{holisticgnn, barad_dur}, or smart storage devices~\cite{cal_nvm_gnn} achieves high performance but requires specialized hardware unavailable to most research groups.
\thiswork targets commodity systems with NVMe SSDs through software optimization.

\noindent\textbf{Storage-Based Training Systems.}
Training large models with storage has become prevalent.
Large language model training systems~\cite{zero_infinity} partition optimizer states, gradients, and parameters across memory hierarchy including NVMe storage.
FlexGen~\cite{flexgen} enables large model inference through offloading and scheduling.
However, these LLM techniques fundamentally differ from GNN requirements.
LLM activations exhibit sequential layer dependencies enabling straightforward layer-by-layer management, while GNN layers exhibit graph-structured dependencies requiring gathering from multiple partitions based on topology.

Mini-batch GNN systems often employ storage for initial feature management.
Ginex~\cite{ginex} introduces unified storage-memory-GPU hierarchy with feature caching.
DiskGNN~\cite{diskgnn} employs pre-sampling and IO-aware scheduling.
GNNDrive~\cite{gnndrive} optimizes feature storage layouts.
MariusGNN~\cite{marius} loads valid features with two-level partitioning.
Helios~\cite{helios} enables direct GPU access to storage.
These systems manage only initial features, as sampled subgraphs fit in GPU memory.
The computational pattern involves: (1) sampling subgraphs, (2) fetching initial features, (3) performing forward-backward in GPU memory with all intermediate activations resident.
Each mini-batch operates independently with no inter-batch dependencies for intermediate activations.
On the other hand, full-graph training requires simultaneous management of all intermediate activations/gradients for all vertices across all layers with inter-partition dependencies spanning the entire graph, creating fundamentally different storage access patterns.
Additional subgraph training frameworks~\cite{graphsage, graphsaint, distdgl, distdglv2, salient, salient_pp, gnnlab, dorylus, p3, pagraph, legion} address memory constraints through sampling~\cite{aligraph, adaptive_sampling_gcn, giant_graph} but introduce input information loss.

\noindent\textbf{Activation Management.}
Checkpointing trades computation for memory via recomputation~\cite{gradient_checkpoint, batchnorm_checkpoint, checkfreq, justintime_checkpointing, distributed_checkpoint, deepfreeze, check_n_run}, storing subset of activations and reconstructing others during backpropagation.
Prior GNN checkpointing~\cite{understanding_gnn_checkpointing, hongtu, dynamic_gnn_checkpointing, federated_gnn_checkpointing, psgraph} applies checkpointing to GNN training.
HongTu~\cite{hongtu} stores snapshots of offloaded partitions but suffers from massive redundancy.
Redundancy-free computation graphs~\cite{redundancy_free_gnn} address computational redundancy in message passing operations, whereas \thiswork targets storage redundancy elimination.
\thiswork introduces regathering for gradient computation, regenerating gathered activations on-demand from compact partition outputs, reducing storage I/O.

Alternative memory optimization strategies include pruning~\cite{pruning, depgraph, filter_pruning, channel_pruning, movement_pruning, group_pruning}, quantization~\cite{ait, qimera, binaryconnect, fixed_quantization, dorefa_net, quantization_cnn, qbert, exact}, and memory-efficient backpropagation~\cite{bptt, revnet, approximate_activation}.
These techniques complement \thiswork by reducing model requirements but do not address memory scalability challenges from intermediate activations proportional to graph size and hidden dimensions.

\noindent\textbf{Graph Partitioning.}
Partitioning is widely adopted for distributed graph processing~\cite{neugraph, hongtu, betty, metis, adagl, distributed_multigpu, g-miner, agl, cagnet, bgl, neutronstar, g3}.
METIS~\cite{metis} employs multi-level framework with coarsening, partitioning, and refinement phases~\cite{mongoose, coarsening_schemes, aggresive_coarsening, k-way-hypergraph-partitioning}, producing high-quality partitions minimizing edge cuts.
Many distributed GNN frameworks~\cite{pipegcn, sancus, distdgl, distdglv2, bgl, g3, neutronstar} employ METIS for minimizing communication cost.
G3~\cite{g3} proposes iterative METIS-based partitioning enhancing three-dimensional parallelism.
However, METIS requires substantial memory---measurements indicate 4.8$\times$--13.8$\times$ input graph size~\cite{metis_memory, mtmetis}.
Alternative frameworks~\cite{adagl, distributed_multigpu, g-miner} attempt load balancing but demand large memory.
Streaming algorithms~\cite{streaming_metis, streaming_partitioning} and online partitioning~\cite{fennel} reduce memory requirements through single-pass processing.
Label propagation methods~\cite{multilevel-k-way, near-linear-time} and scalable frameworks~\cite{spinner} offer alternative approaches.
These methods target distributed systems or assume sufficient memory for loading entire graph structure.
\thiswork introduces lightweight partitioning requiring only small working set memory, enabling partitioning in memory-limited environments where METIS cannot execute.

\end{document}